\documentclass[twocolumn,prb,amsmath,amssymb]{revtex4}

\usepackage{graphics}
\usepackage{graphicx}
\usepackage{dcolumn}
\usepackage{bm}

\def\al{\alpha}

\def\eps{\epsilon}

\def\veps{\varepsilon}

\def\om{\omega}

\def\be{\begin{equation}}

\def\ee{\end{equation}}

\def\bea{\begin{eqnarray}}

\def\eea{\end{eqnarray}}

\def\bc{\begin{center}}

\def\ec{\end{center}}

\def\ra{\rightarrow}

\def\ov{\over}

\def\nonum{\nonumber}

\begin{document}

\title{Variable-range hopping in 2D quasi-1D electronic systems}


\author{Sofian Teber}
\email[E-mail: ]{steber@ictp.trieste.it}

\affiliation{The Abdus Salam ICTP, Strada Costiera 11, 34014, Trieste, Italy}

\date{\today}

\begin{abstract}
A semi-phenomenological theory of variable-range hopping (VRH) is developed for two-dimensional (2D) quasi-one-dimensional (quasi-1D) systems such as arrays of quantum wires in the Wigner crystal regime. The theory follows the phenomenology of Efros, Mott and Shklovskii allied with microscopic arguments. We first derive the Coulomb gap in the single-particle density of states, $g(\varepsilon)$, where $\varepsilon$ is the energy of the charge excitation. We then derive the main exponential dependence of the electron conductivity in the linear (L), {\it i.e.} $\sigma(T) \sim \exp[-(T_L/T)^{\gamma_L}]$, and current in the non-linear (NL), {\it i.e.} $j({\mathcal E}) \sim \exp[-({\mathcal E}_{NL} / \mathcal{E})^{\gamma_{NL}}]$, response regimes (${\mathcal E}$ is the applied electric field). Due to the strong anisotropy of the system and its peculiar dielectric properties we show that unusual, with respect to known results, Coulomb gaps open followed by unusual VRH laws, {\it i.e.} with respect to the disorder-dependence of $T_L$ and ${\mathcal E}_{NL}$ and the values of $\gamma_L$ and $\gamma_{NL}$.
\end{abstract}

\maketitle


\section{Introduction}

\begin{figure}
\includegraphics[width=6cm,height=4cm]{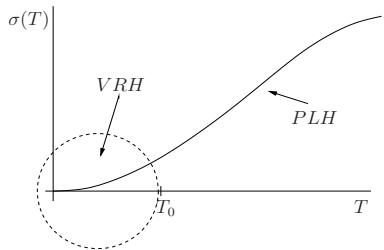}
\vspace{0.1in} \caption{d.c. conductivity as a function of
temperature for a quasi-one dimensional system coupled to
three-dimensional phonons. In the range of temperatures displayed,
phonons  provide a delocalization mechanism for the charge
excitations of the system. Power-law hopping (PLH) dominates at temperatures $T \gg T_0$
and variable-range hopping (VRH) dominates at $T \ll T_0$ ($T_0 \propto 1 /
\tau_{el}$ where $\tau_{el}$ is the elastic scattering time on the static defects). \label{sigma_T}}
\end{figure}

Issues related to transport in low-dimensional electronic systems
are challenging, the main difficulties being to take into account of
disorder or interactions and in some cases of the interplay between them. As known
from Mott and Towse~[\onlinecite{MT}], for strictly one-dimensional (1D)
systems, any disorder leads to localized states. Such a statement
implies that localization and transport properties of
1D Anderson insulators may be tackled perturbatively
by considering the limit of a weak disorder (so called Gaussian
disorder where the density of impurities $N \ra \infty$, their
strength $W \ra 0$ while $NW^2$ is constant). For such a weak
disorder, and in the absence of electron-electron and
electron-phonon interactions, Berezinskii~[\onlinecite{Berezinskii}] has
confirmed the statement of Mott and showed that the a.c.
conductivity is given by: $\sigma_{ac}(\om) \sim \om^2 \log^2
\om$. Subsequently, his approach has been extended by Gogolin,
Mel'nikov and Rashba (GMR)~[\onlinecite{GMR}] to the case where an
electron-phonon coupling is present. They have shown that three-dimensional (3D)
phonons provide a delocalization mechanism for the electrons at
temperatures low enough that the scattering is mainly elastic,
{\it e.g.} $\tau_{in}(T) \gg \tau_{el}$ where $\tau_{in}(T)$ is
the inelastic phonon scattering time and $\tau_{el}$ is the
elastic scattering time on the static defects. This delocalization
mechanism leads to a {\it power-law hopping} for the d.c.
conductivity, {\it e.g.} $\sigma_0(T) \sim T^3$, where the power
comes from the phonon-scattering time. It should be noted that
their arguments are not valid at the lowest temperatures; in
particular, they hold only above $T_0 \propto 1/\tau_{el}$, here and below: $\hbar =1$ unless specified.
Subsequent studies aimed at exploring the effect of
electron-electron interactions in 1D Anderson insulators. In this
respect, it has been shown~[\onlinecite{GS}] that, in a Luttinger
liquid, the Gaussian disorder is strongly renormalized by
interactions. For repulsive interactions, each impurity becomes
effectively strong. The 1D disordered interacting
system is then equivalent~[\onlinecite{KF}] to an ensemble of weak links
where impurities act as wire breakers. As a consequence, the
power-law hopping laws acquire a non-universal exponent,
{\it e.g.} interaction-parameter dependent. More recently~[\onlinecite{GMP}],
the low-temperature situation where the coupling to phonons is
absent and electron-electron interactions dominate has been
addressed. The modern notion of dephasing, due to these
electron-electron interactions, has been considered as the
delocalization mechanism of the electrons. It leads to a power-law
hopping regime, reminiscent of the results of GMR, followed by a
drastic suppression of the conductivity~[\onlinecite{BAA}]. Less explored, from microscopic techniques,
is the low-temperature regime: $T \ll T_0 \propto 1/\tau_{el}$, with both
electron-electron and electron-phonon couplings. In this case, it
is generally believed that the transport is of the {\it
variable-range hopping} (VRH) type. In systems with 3D phonons, that we shall be concerned with in the rest of this manuscript, semi-phenomenological
arguments by Mott~[\onlinecite{Mott_book}] suggest that the VRH laws are characterized by stretched exponentials and read:
\be
\label{L_VRH}
\sigma(T) = \sigma_0(T) \exp[-(T_0 / T)^\gamma],
\ee
where $\gamma \leq 1$. In 1D, $T_0 \propto 1/\nu \xi$, where the
single-particle density of states $\nu \propto 1 / v_F$ and the
localization length $ \xi \propto v_F \tau_{el}$. Hence, $T_0 \propto
1/\tau_{el}$, and the exponential dependence manifests at $T \ll
T_0 \propto 1/\tau_{el}$. In the opposite case, the temperature
dependence of the conductivity arises mainly from the pre-factor:
$\sigma_0(T)$, which depends on the electron-phonon coupling, and
leads to power-law hopping, see Fig.~\ref{sigma_T} for a summary
of regimes. The success of Mott's arguments came from their wide
experimental confirmation for a great variety of disordered
systems, {\it i.e.} it has been found experimentally that $\gamma
\approx 0.25$ in isotropic 3D and $0.33$ in isotropic 2D systems,
in accordance with the theory. Later, Efros and Shklovskii (ES), see Ref.~[\onlinecite{ES_book}] for a review,
extended these arguments to the Mott-Anderson insulators where,
besides strong disorder, the long-range Coulomb interaction is
present. Once again, their results were confirmed by a great
variety of experiments in the field of doped semiconductors where
the exponent $\gamma \approx 0.5$ in all dimensions~\cite{Note:1D} as predicted by ES.

In the present study, we focus on the low-temperature transport
properties of strongly disordered and interacting two-dimensional (2D)
quasi-1D electron systems. The latter consist of a 2D periodic lattice of parallel wires and therefore display markedly anisotropic properties with relations
to the 1D world, see Refs.~[\onlinecite{TG}] and [\onlinecite{Nersesyan}] for reviews on 1D physics. We assume that impurities are point-like and act as wire breakers. Interactions are long-ranged and lead to a large parameter: $r_s = U_H / \veps_F$, corresponding to the ratio of the Coulomb energy-scale: $U_H = e^2 / \kappa a$ ($\kappa$ is the dielectric constant of the host lattice in which the 2D system is embedded), to the kinetic energy scale given by the Fermi energy: $\veps_F$. In the field of charge-density waves (CDW)~\cite{Note:2DCDW}, such quasi-1D Mott insulators are referred to as $4k_F$ CDWs. This $4k_F$ modulation corresponds to a
space periodicity of the charge-density along the wires,
$\lambda_{4k_F}=2 \pi /4k_F$, equal to the average distance between
electrons $a$; such systems therefore display a strongly
correlated state of the Wigner crystal type. For our purposes,
such systems include man-made atomic or molecular wire arrays embedded
in a semi-conducting matrix~[\onlinecite{Himpsel}] and mono-layers of CDWs~[\onlinecite{Gruner_book, Organics_review,ECRYS}]. For these quasi-1D Mott-Anderson insulators, where the interplay between disorder and interactions is quite non-trivial and out of the reach of any perturbative scheme, we will
naturally~\cite{Note:phenom} follow the semi-phenomenological route of Mott and Efros
and Shklovskii in deriving the low-T ({\it i.e.} $T \ll 1/\tau_{el}$)~\cite{Note:quasi-1D_parameter} transport properties. Our
motivation resides in the existing experimental literature on quasi-1D
systems where VRH laws have been reported quite extensively, {\it
{\it e.g.}} in polymers~[\onlinecite{Joo_98,Aleshin_97}].

Eq.~(\ref{L_VRH}) corresponds to {\it linear} variable-range hopping
laws such that the current $j$ is proportional to the applied
electric field $\mathcal{E}$, $j = \sigma(T) \mathcal{E}$. Upon
increasing the applied electric field, such laws cross-over to
{\it non-linear} VRH laws which read:
\be \label{NL_VRH} j(\mathcal{E}) \sim \exp[-(\mathcal{E}_0 /
\mathcal{E})^\gamma], \ee
where the exponent $\gamma \leq 1$. In the frame of disordered
semiconductors in high electric fields such laws, with an exponent
$\gamma = 1/4$, are known from Mott and Shklovskii. In quasi-1D systems
they are observed, {\it e.g.} for bronzes~[\onlinecite{ZKM}].
These laws hold in the low-T limit, $T \ll 1/\tau_{el}$, up to the threshold field for global sliding of
the pinned electronic structure, see Refs.~[\onlinecite{Blatter}] and [\onlinecite{BraNat}] for reviews on pinning. Of course, for the exponential
character to manifest the electric field should not be too high,
{\it e.g.} in 1D: $\mathcal{E} \ll \mathcal{E}_0 \propto 1 / e v_F
\tau_{el}^2$. For a given linear VRH law, it will however manifest
for: $\mathcal{E} > \mathcal{E}_c$ where the crossover field
reads, in any dimension: $\mathcal{E}_c = T \mathcal{E}_0 / T_0
\propto T / e \xi$, where $e$ is the unit charge and $\xi$ is the
localization length. In summary, we shall focus on the following
two regimes: the linear response regime of Eq.~(\ref{L_VRH}):
\be \label{constraint_linear}
T \ll T_0 \approx 1/\tau_{el}~~
{\mathrm{and}}~~ \mathcal{E} \ll T / e \xi,
\ee
and the non-linear response regime of Eq.~(\ref{NL_VRH}):
\be \label{constraint_non_linear}
T \ll T_0 \approx 1/\tau_{el}~~
{\mathrm{and}}~~ T / e \xi \ll \mathcal{E} \ll \mathcal{E}_0.
\ee
Some of the recent theoretical literature on VRH has
been devoted to 1D and 3D quasi-1D
systems in the linear regime, {\it cf.} Ref.~[\onlinecite{FTS}]. To the
knowledge of the author a theory of electron transport in
2D quasi-1D systems has not been developed in either the linear or non-linear regimes. Our specific task will be to compute the main exponential dependence of the
electron current as a function of temperature, electric
field and impurity concentration. We will closely follow the arguments of Ref.~\onlinecite{FTS}.

Our results display two characteristic features: {\it a
non-monotonous dependence of the current as a function of disorder}
and {\it a highly non-universal exponent $\gamma$}, {\it i.e.}
implicitly interaction- and disorder-dependent. Both statements
will be proved in the following sections. They are generic of {\it incommensurate quasi-1D
Mott-Anderson insulators} and are absent in the usual doped
semiconductors or even in the other (hypothetical) collective
structure of interest: the pinned Wigner crystal. The reader interested
more in our results than the details of the derivations may refer directly to Sec.~\ref{conclusion}.

The theory will be developed as follows. In
Sec.~\ref{dielectrics} we describe the $4k_F$ quasi-1D systems in a
semi-classical way, which is justified due to the large parameter
$r_s$. This will unable us to determine the dielectric properties
of the system. With the help of these results and the ES phenomenology we derive, in Sec.~\ref{coulombgap}, the single-particle density of states (DOS) of charge excitations and show how it is affected by the long-range interaction potential among charge excitations (Coulomb gap). In Sec.~\ref{tunneling}
we focus on determining the localization length with the
help of microscopic models related to single-impurity tunneling
of the charge excitations. In Secs.~\ref{linearVRH} and \ref{nonlinearVRH}
the VRH laws, in the linear and non-linear response regimes, respectively, are derived with the help of the Efros, Mott and Shklovskii phenomenology and the results of Sec.~\ref{coulombgap} and Sec.~\ref{tunneling}. In Sec.~\ref{conclusion} the conclusion is given.


\section{Dielectric properties}
\label{dielectrics}

In this Section we determine the dielectric properties of disordered 2D quasi-1D
systems. The latter will be used in subsequent sections to determine the VRH laws.

\subsubsection{The model}

Following the Introduction, we introduce our basic model of a
strongly pinned 2D quasi-1D system in a $4k_F-$CDW state.  Along
each wire, such a quasi-1D system is characterized by a modulation of
the density:
\be
\rho(x) = \rho_0 + \rho_1 \cos(Qx+\varphi) + (1 /2\pi) \partial_x \varphi,
\nonum
\ee
where $\rho_0$ is the unperturbed density, $Q = 4k_F$ is the
modulation of the wave, $\varphi$ is the phase of the (assumed
incommensurate) CDW and the last term describes long-distance
deformations. In electronic systems, a $4k_F$ modulation implies
that the wavelength of the CDW is of the order of the average
spacing $a$ between electrons. It is realized in systems where the
long-range Coulomb interaction is present, see Ref.~[\onlinecite{TG}] for a review on such results in the 1D case. In quasi-1D systems, where the Coulomb interaction may be screened by neighboring chains, the $4k_F-$CDW is also detected~[\onlinecite{Organics_review}]. This reflects the
large $r_s$ nature of such systems. It implies that one may
neglect quantum fluctuations and describe these CDWs
semi-classically with the help of the phase-field $\varphi$. This
CDW phase, $\varphi$, is related to the order-parameter
describing the condensate which reads: $\Delta({\bf r}) = |\Delta|
\exp[i \varphi({\bf r})]$, where $\bf{r}$ is a two-dimensional
coordinate, and $|\Delta|$ the amplitude which is assumed to be
frozen as we consider low temperatures. Moreover, the system has
charge invariance, {\it i.e.} $\varphi \ra \varphi +2 \pi$,
implying the existence of a Fr\"ohlich mode, {\it i.e.} the
sliding of the electron crystal, in accordance with the fact that
the system is incommensurate.

We focus first, in the frame of this semi-classical theory, on
static properties of the system: the determination of the
structure of the electron system in the presence of a single
strong impurity and of it's charge excitations as well as the electrostatic potential
between such excitations [the dynamics will be considered later, {\it i.e.} the
effect of quantum fluctuations, when dealing with the tunneling
of these charge carriers through the impurities]. Adding then a
single strong impurity at the origin of the 2D system, the
Hamiltonian consists of three parts:
\begin{eqnarray}
\label{H2D}
H = H_0 + H_D + H_I,
\end{eqnarray}
where $H_0$ is the elastic part and reads:
\begin{eqnarray}
\label{H2D_0}
H_0 = \int{{d^{2} {\bf r} }}~{Y \over 2}~
[ \left( \partial_x \varphi \right)^2 + \alpha \left(\partial_y \varphi \right)^2],
\end{eqnarray}
where $Y = Y_x$ is the bulk-modulus along the chains
and the anisotropy parameter reads: $\alpha =
Y_y / Y_x \ll 1 $ ($x-$ is the direction along the wires and $y-$ is the direction
perpendicular to them). In Eq.~(\ref{H2D_0}), the phase field
$\varphi$ may be interpreted as a scalar displacement
field along the wires, $u_x$, as in conventional elasticity
theory. The displacement and phase fields are then related by:
$\varphi = - (2 \pi /a) u_x$. It follows naturally that the first
term, in Eq.~(\ref{H2D_0}), describes the compression energy
along each wire, $Y$ being the bulk modulus and the second term
corresponds to shear elasticity due to inter-wire interactions,
$Y_y$ being the shear modulus. It should be noticed that the
inter-wire interaction term is defined at the level of the order
parameter with the help of: $\Delta^*_n \Delta_m + \mathrm{c.~c.}=
|\Delta|^2 \cos(\varphi_n - \varphi_m)$, where $n$ and $m$ index
neighboring wires. Shear elasticity is then derived by expanding
the cosine, at low temperatures, and going to the continuum limit
perpendicular to the wires.

The second term in Eq.~(\ref{H2D}) describes the effect of a
strong impurity at the origin and reads:
\begin{eqnarray}
\label{H2D_D} H_D = \int{{d^{2} {\bf r} }} ~[ W_f \partial_x
\varphi \delta({\bf r}) - W_b \cos(\varphi) \delta({\bf r})],
\end{eqnarray}
where the first term corresponds to the forward scattering on the
impurity ({\it i.e.} the coupling of the point-like impurity
potential to the long-distance part, $\propto \partial_x \varphi$,
of the CDW density) and has a strength $W_f$. The second term
corresponds to the backscattering on the impurity ({\it i.e.} the
coupling of the point-like impurity potential to the oscillating,
$4k_F$, part of the CDW density) with a strength $W_b$.

Finally, the third term in Eq.~(\ref{H2D}) contains the long-range
Coulomb field and reads:
\begin{eqnarray}
\label{H2D_I} H_I = \int{{d^{2} {\bf r} }} ~[ U \delta({\bf r}) +
{1 \over b^2} \partial_x \varphi U({\bf r})] - {1 \over 8 \pi e^2}
\int{{d^{3} {\bf r} }}({\bf \nabla }U)^2],
\end{eqnarray}
where $U$ is the Coulomb field. The first term corresponds to a
point-like test charge for the Coulomb field $U$. The second to
the coupling of the long-range Coulomb potential to the
long-distance part of the density ($\propto \partial_x \varphi$)
and the last term corresponds to the energy of the Coulomb field.
It should also be noticed that, in the model of Eq.~(\ref{H2D_I}),
the Coulomb interaction is the real one, {\it i.e.} the
three-dimensional one.

There is a basic non-trivial length-scale in
Eq.~(\ref{H2D_0}) which is proper to quasi-1D systems and will bring significant opportunities to go beyond strictly 1D physics all along this manuscript. This
is the length of a $2\pi-$soliton: $l_s$, and emerges due to the
fact that in quasi-1D systems shear and elasticity are
coupled~\cite{Note:ls}. This can be understood with the help of
the following arguments. An impurity will enforce a deformation of
the CDW in its vicinity. In the absence of long-range Coulomb
interaction, a deformation $\delta \varphi$ along a distance
$\delta x$ along the defected wire, has an elastic energy: $E =
\delta x~ b^2 [ Y \delta \varphi^2 / \delta x^2 + Y_\bot  \delta
\varphi^2 / b^2]$, where we have assumed that the distance between
neighboring wires is $b$. Minimizing this elastic energy with
respect to $\delta x$ and defining the optimal $\delta x$ as the
length $l_s$, we find that:
\begin{equation}
\label{l_s}
l_s = {b \over \sqrt{\alpha}}.
\end{equation}
In the case where $\delta \varphi = 2 \pi$, the deformation
becomes the plastic one, as a whole period of the density-wave is
affected by the impurity. Hence the name of {\it soliton} length, see Refs.~[\onlinecite{BK}] and~[\onlinecite{Lu}] for reviews on soliton physics in condensed matter. Notice also that for decoupled wires, {\it i.e.} strictly 1D
systems where $\alpha \rightarrow 0$, $l_s$ is infinite. This
implies that the whole wire adjusts to the deformation at the
origin. In the case of~ $0 < \alpha < 1$, inter-wire interactions
enforce the same phase $[2 \pi]$ between neighboring wires, {\it
beyond} the length $l_s$, on both sides of the impurity. In what
follows, the length $l_s$ will be used to distinguish between the
low-impurity concentration limit, where the average distance
between impurities along a wire, $l$, is larger than $l_s$, $l \gg
l_s$ and the opposite large-impurity concentration limit, $l \ll
l_s$ (the density of impurities is given by: $N = 1 / l b^2$).

\subsubsection{The case of small impurity concentration}

We focus on the low-impurity concentration case, $l=1/Nb^2 \gg
l_s$ (recall that $l$ is the average distance between impurities along a wire), {\it cf.} the one-impurity model of Eq.~(\ref{H2D}):
\be
N \ll N_s = {1 \ov l_s b} = {\sqrt{\alpha} \ov b^2}.
\label{N_S}
\ee
Each impurity is strong and the backscattering term in the
Hamiltonian describes the pinning of the system at distances of
the order of the average distance between impurities, $l$. On the
other hand the forward scattering term describes softer
deformations of the system away from the impurities [{\it cf.} the
dipole distortions that we will introduce below]. We focus on the
latter~\cite{Note:fscatt} and set: $W_b = 0$, assuming that the system is strongly
pinned. By varying the functional of Eq.~(\ref{H2D}) with respect
to the phase field and the Coulomb field, $\delta H/ \delta
\varphi=0$ and $\delta H/ \delta U=0$ and going to Fourier space,
one gets a system of two Poisson equations where both fields
screen each other:
\begin{subequations}
\label{EMq_2D}
\begin{equation}
\varphi({\bf q}) = { i q_x \over {\tilde q}^2 \veps_2({\bf q})} \left[{W_f \over Y} +
{b \over r_Dq} \right],
\label{phi_q_2D}
\end{equation}
\begin{equation}
U({\bf q}) = {2 \pi e^2 \over q \veps_2({\bf q})} \left[ 1 -
{W_f \over Y b}{q_x^2 \over {\tilde q}^2} \right],
\label{U_q_2D}
\end{equation}
\end{subequations}
where ${\tilde q}^2 = q_x^2 + \alpha q_y^2$, $q^2 = q_x^2 +
q_y^2$ and $r_D^{-1} = 2\pi e^2 / Y b^2 \approx b^{-1}$ is the inverse
screening length in the metallic phase. The 2D dielectric constant
with the help of which the long-range Coulomb and elastic fields
screen each-other reads:
\begin{equation}
\label{eps_2D} \eps_2({\bf q}) = \kappa \left[ 1 + {q_x^2
\over r_D q {\tilde q}^2} \right],
\end{equation}
which displays the anisotropic metallic-like (in the sense of sliding of the collective electron
structure) screening in the quasi-1D system. As can be seen from Eqs.~(\ref{phi_q_2D})
and~(\ref{eps_2D}), in 2D, the effect of the Coulomb interaction
is to shift ${\tilde q}^2$ to ${\tilde q}^2 \veps_2({\bf q})=
{\tilde q}^2 + q_x^2 / r_D q$. This shift can be included in the
bulk modulus and corresponds to the incompressibility of the
crystal as a whole. As we are interested in physical properties at
large distances along the chains $q_x \ll \sqrt{\alpha} q_y$ or $x
\gg y/ \sqrt{\alpha}$, we see that the Coulomb interaction brings
into play the following sector:
\begin{equation}
\label{sector_2D}
q_x^2 \ll \alpha r_D q_y^3 \qquad y \ll ( \alpha r_D x^2 )^{1/3},
\end{equation}
which plays a dominant role as we show now.

The inverse Fourier transform of Eq.~(\ref{EMq_2D}) yields:
\begin{subequations}
\label{EMr_2D}
\bea
&&\varphi({\bf r}) =  {-b~ {\mathrm{sgn}}(x)
\over 6\pi (\sqrt{\alpha r_D} |x| )^{2/3}} \left[ 3\Gamma({5 \over
3}) - 2 \Gamma({2 \over 3},{\sqrt{\alpha r_D} |x| \over y^{3/2}})
\right],~~~~
\label{phi_r_2D} \\
&&U({\bf r}) =  {e^2 \over  \kappa |x|} \left[ 1 -
\exp \left(- {\sqrt{\alpha r_D} |x| \over y^{3/2}} \right) \right],~~~~
\label{U_r_2D}
\eea
\end{subequations}
where $\Gamma(x)$ and $\Gamma(x,y)$ are the complete and incomplete gamma functions, respectively. The phase field of Eq.~(\ref{phi_r_2D}) is an odd function of the
coordinate along the wire and at long distances along the wire,
{\it i.e.} for $y \ll ( \alpha r_D x^2 )^{1/3}$, reads: $\varphi({\bf
r}) \propto - (l_s / x)^{2/3}$, {\it cf.} Eq.~(\ref{phi_2D_IN}) below.
This implies that, at such distances, the impurity is surrounded
by {\it dipole distortions} of length $l_s$. Crucially, we identify these dipole
distortions as the charge excitations of the system.

These charge excitations interact with a Coulomb potential which deviates
from the usual 3D one, due to the anisotropic screening, {\it cf.}
Eq.~(\ref{U_r_2D}). More generally, Eq.~(\ref{U_q_2D}) shows that the electrostatic
potential changes sign along the cone: $y = \pm \sqrt{\alpha} |x| /
(W_f/Yb -1)$. This is related to the dipole nature of the phase
deformations. In particular, outside this cone, the potential
reads: $U \propto -W_f r_D /r^3$, so that charge deformations of the
same sign attract each-other via a dipole potential.
On the other hand, within this cone
(which is the sector of validity of Eq.~(\ref{U_r_2D})) the
potential is repulsive and independent of $W_f$. This implies that, if they were not bound to the impurities (which they actually originate from), such charge excitations would form domain walls (our arguments have their roots in Ref.~[\onlinecite{BM}]). In the following we return on the sector within the cone: $y < \pm \sqrt{\alpha} |x| / (W_f/Yb -1)$, where the potential is repulsive and
Eqs.~(\ref{EMr_2D}) holds.

We summarize this sub-section by giving the asymptotic expressions of
Eqs.~(\ref{EMr_2D}), on one hand close to the chains, {\it i.e.} in the sector of Eq.~(\ref{sector_2D}), $y \ll ( \alpha r_D x^2 )^{1/3}$, where it reads :
\begin{subequations}
\label{EMr_2D_IN}
\bea
&&\varphi({\bf r}) = - {\mathrm{sgn}}(x)~\left({l_s \over 2 \pi |x| } \right)^{2/3},
\label{phi_2D_IN} \\
&&U({\bf r}) =  {e^2 \over  \kappa |x|},
\label{U_2D_IN}
\eea
\end{subequations}
and on the other hand further away from the chains, {\it i.e.} $y \gg (
\alpha r_D x^2 )^{1/3}$, where it reads:
\begin{subequations}
\label{EMr_2D_OUT}
\bea
&&\varphi({\bf r}) =  - {b~{\mathrm{sgn}}(x) \over 2 \pi |y|},
\label{phi_2D_OUT} \\
&&U({\bf r}) =  {e^2 \sqrt{\alpha r_D} \over \kappa y^{3/2}}.
\label{U_2D_OUT}
\eea
\end{subequations}
In the limit of vanishing inter-wire coupling: $\alpha \rightarrow 0$,
the sector defined by Eq.~(\ref{sector_2D}) vanishes. This
corresponds effectively to a crossover from a low impurity density
regime, $N \ll N_s = {1 / l_s b^2} = {\sqrt{\alpha} / b^3}$, to a
large impurity density regime, $N \gg N_s$, that we consider next.

\subsubsection{The case of large impurity concentration}

Formally, this case, as defined in the previous section, requires
that $l \ll l_s$ (recall that $l$ is the average distance between impurities along a wire) or, in terms of impurity concentration, that:
\begin{equation}
\label{N_L} N \gg N_s={\sqrt{\alpha} \over b^2},
\end{equation}
where $\alpha$ gives the dimensionless strength of inter-chain
interactions. Large impurity concentration is therefore equivalent
to vanishing inter-chain couplings. The system is then equivalent
to an ensemble of metallic (in the sense of sliding) segments along the wires, of average length $l$. These segments are decoupled elastically but still coupled by the long-range Coulomb potential. In the literature,
such a regime is sometimes referred to as a model of interrupted metallic
strands~\cite{RB}. Following the previous paragraph, our goal here is to
determine the dielectric properties of such a phase.

From Eqs.~(\ref{EMq_2D}), the electrostatic potential of a charge
carrier at the origin of a {\it pure} system ($W_f = W_b =0$)
reads:
\begin{equation}
\label{U_2D}
U({\bf q}) = {2 \pi e^2  \over q \eps_2({\bf q})},
\end{equation}
where the dielectric constant is given by Eq.~(\ref{eps_2D}) and $q^2 = q_x^2 + q_y^2$. In the previous paragraph we have focused on the effect of
deformations due to the forward scattering term ($W_f$) in a system of
dilute strong pinning centers. Increasing the impurity
concentration we need to take into account the average effect of
these pinning center. This is done according to the prescription:
$ {\tilde q}^2 \rightarrow {\tilde q}^2 + L_x^{-2}$ in
Eq.~(\ref{eps_2D}) for the dielectric constant (recall that ${\tilde q}^2 = q_x^2 + \alpha q_y^2$ arises from the elastic part of the energy) where $L_x \propto l$ is the pinning length due to the backscattering on the impurity, see Refs.~\onlinecite{Blatter} and \onlinecite{BraNat} for reviews on pinning. This amounts to introduce the effect of backscattering on the impurity as a commensurability term
explicitly breaking the translational invariance of the system.
The dielectric constant therefore becomes:
\begin{equation}
\label{eps_L_2D}
\eps_2({\bf q}) = \kappa \left[ 1 + {q_x^2 \over r_D q  ({\tilde q}^2 + L_x^{-2})} \right].
\end{equation}
In the clean limit, $L_x \propto l \gg l_s$, we may assume that $L_x/l_s \ra \infty$, and recover the results of the previous subsection. In the dirty limit, $L_x \propto l \ll l_s$, we may assume that $l_s/L_x \ra \infty$ so that $\alpha \ra 0$ and ${\tilde q}^2 = q_x^2 + \alpha q_y^2 \ra q_x^2$. In this regime, the Coulomb potential reads:
\be
U({\bf r}) =
{2 \pi e^2 \ov \kappa}~
\int {{d{\bf q} \ov (2\pi)^2}}~{e^{i {\bf q}.{\bf r}} \ov q + {q_x^2 \ov r_D~(q_x^2 + L_x^{-2})}}.
\label{U_2D_FT}
\ee
The asymptotics of Eq.~(\ref{U_2D_FT}) then read:
\begin{subequations}
\label{phi_Keldysh}
\bea
U({\bf r}) =&& \frac{e^2}{\kappa |x|} \left( 1 - \exp \left(-
\frac{|x|}{\sqrt{y b \kappa_x / \kappa}} \right) \right),
\nonum \\
&& ~~~~~~~ L_x \ll x \ll L_x^2/2b,~~ y \ll L_x^2/2b,
\label{KAD1} \\
U({\bf r}) =&& \frac{e^2}{\kappa y},~~L_x \ll x \ll L_x^2/2b,~~ y \gg L_x^2/2b,~~~
\label{KAD2} \\
U({\bf r}) =&& \frac{e^2}{\kappa r}, \quad r \gg L_x^2/2b,
\label{KLD}
\eea
\end{subequations}
where $r=\sqrt{x^2+y^2}$ and $\kappa_x$ is the longitudinal (along the wires) dielectric constant:
\be
\kappa_x = \kappa (L_x/r_D)^2,
\label{kappax}
\ee
up to a numerical constant of the order of unity, with: $r_D \approx b$, where $b$ is the inter-wire distance and the pinning length along the wires: $L_x \approx l$, is of the order of the average impurity-distance, $l$, along each wire. Notice that the longitudinal dielectric constant, Eq.~(\ref{kappax}), is inversely proportional to the square of the impurity concentration in the system, $N=1/lb$. This longitudinal dielectric constant is large, {\it i.e.} larger than the dielectric constant $\kappa$ of the host in which the 2D system is embedded. This is important and explains the richness of Eqs.~(\ref{phi_Keldysh}). Indeed, Eq.~(\ref{KAD1}) shows that close to the chains ($y \ll l^2/2b$) and for $l \ll x \ll l^2/2b$, field lines prefer to remain in the 2D layer where they are screened by the large longitudinal dielectric constant. This sector is peculiar to the quasi-1D system. At larger distances, $x \gg l^2/2b$, the lines escape to the external media where they are screened only by $\kappa$ and the potential is fully long-ranged. This result is derived by other means in the Appendix.


\section{Single-particle density-of-states and the Coulomb gap}
\label{coulombgap}

We now turn to the determination of the single-particle density of
states, $g(\veps)$, of the localized charge excitation of
energy  $\veps$. This density of states is defined as the
probability density for the excitation to have energy $\veps$.
In the case where $\veps = \veps_i$, the corresponding wave
function is given by:
\begin{equation}
\label{psi}
\psi_i({\bf r}) = \exp[- {x_i \over \xi_x} - {y_i \over \xi_y}],
\end{equation}
where $\xi_x$ is the longitudinal, {\it i.e.} along the wires,
localization length and $\xi_y$ is the transverse one. Moreover,
$x_i$ and $y_i$ locate the position of the localized charge
excitation of energy $\veps_i$. Notice that, in usual
disordered semiconductors, $\psi$ is the single-particle wave function of
the electron (impurities are hydrogenoid atoms) and the
localization length corresponds to the Bohr radius of this
trivial charge excitation. In quasi-1D systems $\psi$
is the single-particle wave function of deformations of the electronic system,
the distortions considered in the previous section.

An ensemble of such localized states with a density $g(\veps)$,
forms a disorder or impurity band. Moreover, keeping in mind that the system is a
Mott insulator, the localized states of energy $\veps$ belong
the lowest Hubbard band of the system. It is on this lowest
"Hubbard impurity-band" that we shall focus now, keeping in mind
that the upper one, distant by the Hubbard: $U_H \sim e^2 / \kappa
a$, is unreachable at the low energies we consider: $T \ll U_H$.

We further assume that {\it this lowest Hubbard impurity band is
partially filled}. As a result, the low-T
transport we consider is the hopping of charge excitations within
this lowest Hubbard impurity band. One may formally define a Fermi
energy for this band of localized states ($\veps_F \equiv 0$)
and a non-zero DOS of charge excitation at this Fermi level: $\nu
\equiv g(0) \not= 0$. As far as transport is concerned, we shall
show below that the non-zero $\nu$, {\it i.e.} the existence of
gapless charge excitations, implies that there is a
low-temperature hopping conductivity of the variable-range type.

In the absence of the long-range Coulomb $g(\veps)$ is
constant, {\it i.e.} $g(\veps) \approx \nu$, within the disorder band
width. We now turn on the long-range Coulomb interaction and
follow Efros and Shklovskii to determine the influence of this
interaction on $g(\veps)$. We provide here some details for the reader which is not familiar with the ES arguments. While hopping from the localized state $i$ of energy $\veps_i$ below the effective Fermi energy of the impurity-band to the state $j$ of energy $\veps_j$ above, the energy of the charge excitation will vary by: $\delta E
= \veps_j - e^2 / \kappa r_{ij} - \veps_i$. In the latter expression,
the long-range Coulomb interaction $- e^2 / \kappa r_{ij}$ between
the charge excitation at $j$ and the hole it has left at $i$ has
been taken into account. By construction: $\delta E \geq 0$, which
implies a {\it depletion} of states around the Fermi energy:
$r_{ij} \geq e^2 / \kappa |\veps_j - \veps_i|$, because of
the long-range nature of the Coulomb interaction. In $d$
dimensions, such states have a spatial density of $n = 1/ r^d$,
where $r \equiv r_{ij}$. This implies that: $n(\veps) \leq
(\kappa |\veps| / e^2)^d$, where $\veps \equiv \veps_j -
\veps_i$. The corresponding single-particle density of states
($g(\veps) = dn(\veps) / d \veps$) then reads:
\be
\label{CG_iso}
g(\veps) = C~( {\kappa \over e^2})^d~|\veps|^{d-1},
\ee
where $C$ is a numerical coefficient of the order of unity,
depending on the dimensionality of the system~\cite{Note:1D}, {\it e.g.} see
Ref.~\onlinecite{Nguen} for the 2D case we are interested in. A crucial feature
of Eq.~(\ref{CG_iso}) is that the DOS vanishes only at the Fermi
energy. Hence, there is still no Coulomb blockade within the
impurity band because of the long-range Coulomb. There is however
a soft Coulomb-gap which affects the transport as will be shown
below. Moreover, the Coulomb gap of Eq.~(\ref{CG_iso}) depends on dimensionality~\cite{Note:1D}: quadratic in the energy of the
charge excitation in 3D ($\propto \veps^2$) and linear in 2D
($\propto |\veps|$), and not on the impurity concentration. The situation is much richer in quasi-1D systems due to their non-trivial dielectric properties.

\subsubsection{The case of large impurity concentration}

We apply the Efros-Shklovskii arguments to the case of the
anisotropic Coulomb interaction characteristic of quasi-1D systems
starting with the large-impurity concentration regime: $N \gg N_s$. The Coulomb
interaction is given by  Eqs.~(\ref{phi_Keldysh}). For each sector
we fix the potential $U=\veps$, determine the equipotentials
$x(\veps)$ and $y(\veps)$ and substitute them in the
density, $n(\veps)$ in order to derive the Coulomb gap:
\begin{equation}
n(\varepsilon) = {1 \over x(\varepsilon)y(\varepsilon)}, ~~~
g(\veps) = {d n(\veps) \ov d \veps}.
\nonum
\end{equation}
We first focus on the large-distance sector of Eq.~(\ref{KLD}). At such distances, $x \gg l^2/2b$, the potential is isotropic. Therefore the usual ES arguments apply, $x(\veps) = y(\veps) = e^2/\kappa \veps$ and $n(\veps) = \kappa^2 \veps^2 / e^4$. The Coulomb gap is therefore linear in the energy of the charge excitation:
\be
g_{ES}(\veps) = C_0~{\kappa^2 \over e^4}~|\veps|, \quad \veps \ll \veps_1,
\label{gES}
\ee
where $C_0$ is a numerical coefficient of the order of unity, $\veps_1$ is a crossover to another Coulomb-gap shape (as explained shortly). As in the usual ES law, Eq.~(\ref{gES}) is independent on the impurity concentration ($C_0$ and $\kappa$ are constants).

\begin{figure}
\includegraphics[width=5cm,height=3cm]{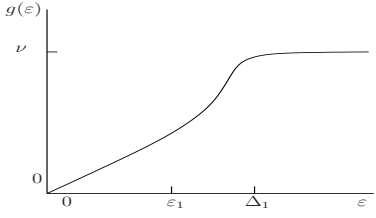}
\caption{ \label{DOS_2D_L} Schematic plot of the DOS of charge excitations of the 2D anisotropic system
with a large concentration of impurities, $N \gg N_s$. The linear Coulomb-gap at low energies
[Eq.~(\ref{gES})] is succeeded by the quadratic Coulomb-gap [Eq.~(\ref{g1})] and finally
by the constant DOS of the disorder band, $\nu$.}
\end{figure}

We then focus on the shorter-distance sector of Eq.~(\ref{KAD1}). At such distances, $l \ll x \ll l^2/2b$ and $y \ll l^2/2b$ (recall that $L_x \propto l$), the potential reads:
\begin{subequations}
\label{parabola}
\bea
&&U({\bf r}) = {e^2 \over \kappa x}, \qquad ~~~~~~~~ y \ll {x^2 \kappa \over b \kappa_x},
\label{parabola_a} \\
&&U({\bf r}) = {e^2 \over \kappa \sqrt{y b \kappa_x / \kappa}}, \quad  y \gg {x^2 \kappa \over b \kappa_x}.
\label{parabola_b}
\eea
\end{subequations}
which follows from Eq.~(\ref{KAD1}). Close to the chains, $y \ll {x^2 \kappa / b \kappa_x}$, and fixing $U \equiv \veps$ yields the following equipotentials: $x(\veps)=e^2/\kappa \veps$ and $y(\veps)=x(\veps)^2 \kappa / a \kappa_x$.
This yields a density of localized states:
\begin{equation}
n(\veps) = {b \kappa^2 \kappa_x \veps^3 \over e^6},
\nonum
\end{equation}
and therefore the Coulomb gap:
\begin{equation}
\label{g1}
g_{1}(\varepsilon) = C_1~{b \kappa^2 \kappa_x \varepsilon^2 \over e^6},
\quad \varepsilon_1 \ll \varepsilon \ll \Delta_1,
\end{equation}
where $C_1$ is a numerical coefficient of the order of
unity. This Coulomb gap is {\it quadratic} for a 2D system and depends on disorder through $\kappa_x$, {\it cf.} Eq.~(\ref{kappax}). Both features are unusual with respect to known results, {\it cf.} Eq.~(\ref{CG_iso}). In Eq.~(\ref{g1}), $\Delta_1$ is the width of the Coulomb gap and $\veps_1$ a crossover energy from Eq.~(\ref{gES}) to Eq.~(\ref{g1}). The upper-bound $\Delta_1$ to the energy-dependence of Eq.~(\ref{g1}) originates from the fact that this Coulomb gap is due to the short-distance part of the Coulomb potential Eq.~(\ref{KAD1}). At shorter-distances, hence higher energies, Eq.~(\ref{g1}) crosses-over to the constant DOS, $\nu$, of the disorder band. Equating $g_{1}(\Delta_1)$ to $\nu$ yields:
\be
\Delta_1 = D_1~\left[ e^6 \nu / b \kappa^2 \kappa_x \right]^{1/2},
\label{Delta_1}
\ee
where $D_1$ is a numerical coefficient of the order of
unity. Eq.~(\ref{Delta_1}) corresponds to the Coulomb-gap width in the large impurity regime.
On the other hand, the crossover energy $\veps_1$ is obtained by equating $g_{ES}(\veps_1)$ to $g_1(\veps_1)$ which, from Eqs.~(\ref{gES}) and (\ref{g1}), yields:
\be
\veps_1 = N_1~{e^2 \ov b \kappa_x},
\label{eps1}
\ee
where $N_1$ is a numerical coefficient of the order of
unity. Notice that both Eqs.~(\ref{Delta_1}) and (\ref{eps1}) have an impurity-dependence through the longitudinal dielectric constant $\kappa_x$ which is given by Eq.~(\ref{kappax}). In the large impurity concentration case, the total Coulomb gap shape is plotted on Fig.~\ref{DOS_2D_L}.

\subsubsection{The case of small impurity concentration}

\begin{figure}
\includegraphics[width=5cm,height=3cm]{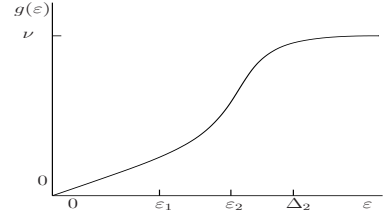}
\caption{ \label{DOS_2D_S} Schematic plot of the DOS of the 2D anisotropic system
with a small concentration of impurities, $N \ll N_s$. The linear Coulomb-gap at low energies [Eq.~(\ref{gES})]
is succeeded by the quadratic Coulomb-gap [Eq.~(\ref{g1})], then by the "2/3-"Coulomb-gap
[Eq.~(\ref{g2})] and finally by the constant DOS of the disorder band, $\nu$.}
\end{figure}

We turn on to the small-impurity concentration case: $N \ll N_s$.
The ES Coulomb gaps found in the case of large impurity concentration, $N \gg N_s$, are still valid in the present case.
Hence, at distances $x$ larger than $l^2/2b$,
interactions open a linear Coulomb gap similar to the one of Eq.~(\ref{gES}).
At smaller distances: $l \ll x \ll l^2/2 b$, corresponding to larger energies, this
linear Coulomb gap crosses over, at $\varepsilon_1$, to the quadratic one defined by Eq.~(\ref{g1}).
Going to higher energies corresponds to distances shorter than the pinning length, $L_x \propto l$. If impurities are sufficiently diluted, {\it i.e.} $l$ is large, we may face the situation where a charge excitation experiences the potential within the pinning area $L_xL_y = \sqrt{\alpha}l^2$, {\it i.e.} the potential of the equivalent pure system given by Eqs.~(\ref{U_2D_IN}) and (\ref{U_2D_OUT}):
\begin{subequations}
\label{phi_Small2}
\bea
&&U({\bf r}) = {e^2 \over \kappa |x|},  \qquad ~~~~~~~~ y \ll [x^2 \alpha r_D]^{1/3}
\label{phi_Small2a} \\
&&U({\bf r}) = {e^2 \over \kappa}
\left[{\alpha r_D \over |y|^3} \right]^{1/2}, \quad y \gg [x^2 \alpha r_D]^{1/3}.
\label{phi_Small2b}
\eea
\end{subequations}
Close to the chains, $y \ll [x^2 \alpha r_D]^{1/3}$, and fixing $U \equiv \veps$ yields the following equipotentials: $x(\veps)=e^2/\kappa \veps$ and $y(\veps)= [x^2(\veps) \alpha r_D]^{1/3}$.
This yields a density of localized states:
\begin{equation}
n(\veps) = {1 \ov (\al r_D)^{1/3}}~\left( \kappa \veps \ov e^2 \right)^{5/3},
\nonum
\end{equation}
and therefore the Coulomb gap:
\begin{equation}
\label{g2}
g_{2}(\varepsilon)= C_2~{ \varepsilon^{2/3} \over (\alpha r_D)^{1/3} (e^2/\kappa)^{5/3} },
\quad \varepsilon_2 \ll \varepsilon \ll \Delta_2,
\end{equation}
where $C_2$ is a numerical coefficient of the order of
unity. In Eq.~(\ref{g2}), the new crossover energy, $\varepsilon_2$, defined as the crossover energy between the quadratic and $2/3-$Coulomb gaps [$g_{1}(\varepsilon_2)=g_{2}(\varepsilon_2)$] reads:
\begin{equation}
\label{eps2}
\varepsilon_2 = N_2~{e^2 \over \kappa b }~\left[ {1 \ov \alpha}
\left( {\kappa_x \ov \kappa} \right)^3 \right]^{1/4},
\end{equation}
where $N_2$ is a numerical coefficient of the order of
unity.
Finally, $\Delta_2$ corresponds to the Coulomb-gap width in the small-impurity concentration case [$g_{2}(\Delta_2)=\nu$]
and reads:
\begin{equation}
\label{Delta2}
\Delta_2 = D_2~\left[ (e^2 / \kappa)^5 \nu^3 \alpha r_D \right]^{1/2},
\end{equation}
where $D_2$ is a numerical coefficient of the order of
unity. These results are summarized on Fig.~\ref{DOS_2D_S} which displays the total DOS of the system in the small impurity concentration regime.


\section{Localization length and tunneling}
\label{tunneling}

We now focus on the localization length associated with the
single-particle wave function of Eq.~(\ref{psi}). The transverse localization length is given by:
$\xi_y < b$, where $b$ is the inter-chain distance. In the following we focus on $\xi_x$.
As we have already said at the level of Eq.~(\ref{psi}), in the case of doped semiconductors, the charge
excitations are simply the electrons and the corresponding
localization length is the Bohr radius of the hydrogenoid wave
function. The later has no impurity dependence. On the other hand,
the situation is much less trivial for quasi-1D
systems, where excitations are non-linear deformations of the
electronic system. These excitations are extended and one may
think that they are localized over the length $\xi_x = l_s$ along the wires, in the clean
limit ($l \gg l_s$) and over the distance $\xi_x = l$ between impurities
in the dirty limit ($l \ll l_s$).
This intuitive argument is correct but it turns out that the F\"ohlich mode as well as the
long-range Coulomb interaction also influence, in a non-trivial
way, this length, as we show now. The reader which does not wish to go through
our microscopic arguments, at least on the first reading, may skip to the next section, where
transport laws are derived and it is shown how the localization lengths enter expressions for the conductivity and current.

\subsubsection{The model}

It is crucial to notice that the localization length is of purely
quantum origin, contrary to the pinning length (Fukuyama-Lee
length or Larkin length depending on the context) which is defined
at the classical level. The derivation of the localization length
requires to consider the tunneling of the charge excitations
through impurities. The most convenient way of dealing which such
processes, which are forbidden at the classical level, is to
extend the semi-classical approach of previous sections to the
Euclidean space by using imaginary time dynamics. In doing so, we
consider the tunneling of a charge excitation along the distance
$x$ of a given wire. The average distance between impurities along
this wire is $l$. Hence, the charge has to tunnel through $x/l$
impurities. Neglecting interferences between these impurities,
their effect, as a first approximation, is additive. The total
action for tunneling along the distance $x$ is therefore given
by: $S_{\mathrm{opt}}(x) = s_{\mathrm{opt}} x / l$, where
$s_{\mathrm{opt}}$ is the action to tunnel through {\it a single}
impurity. The total action $S_{\mathrm{opt}}$ enters the quantum
probability to reach the distant point $x$: $|\psi(x)|^2 \sim
\exp(-S_{\mathrm{opt}}(x))$, where $\psi$ is the
one-(quasi-)particle wave function of Eq.~(\ref{psi}). Therefore:
$x/\xi_x \equiv S_{\mathrm{opt}} = s_{\mathrm{opt}} x / l$ and the
localization length reads:
\begin{equation}
\label{xi_x_def1}
\xi_x = l / s_{\mathrm{opt}}, \quad l \ll l_s.
\end{equation}
This result is valid for the large impurity concentration case. When the concentration of impurity is small charge excitations acquire the length $l_s$ and the localization length becomes:
\begin{equation}
\label{xi_x_def2}
\xi_x = l_s / s_{\mathrm{opt}}, \quad l \gg l_s.
\end{equation}
In our approximation of neglecting interferences between various
impurities, which is reasonable in the strong or individual
pinning regime we consider, the localization length depends only
on a one-impurity tunneling action. Subsequent calculations will
therefore focus on a derivation of this one-impurity optimal
action $s_{\mathrm{opt}}$.

The basic action that we shall consider reads:
\begin{eqnarray}
\label{quantum_action}
s = \int_0^\beta {d\tau} \int {d{\bf r}}~
\{&& {C \over 2} (\partial_\tau \varphi)^2 +
{Y \over 2} \left[ \left( \partial_x \varphi \right)^2
+ \alpha \left( \partial_\bot \varphi \right)^2 \right]
\nonumber \\
&&- W_b \cos(\varphi ) \delta({\bf r}) \},
\end{eqnarray}
where now $\varphi \equiv \varphi(\tau,{\bf r})$, $C$ is the CDW
stiffness, the last term corresponds to the backscattering on the
impurity and $\hbar=1$ unless specified explicitly. At this point
we came up to a systematic way of deriving the localization length
with the help of Eqs.~(\ref{xi_x_def1}), (\ref{xi_x_def2}) and the well-defined model
of Eq.~(\ref{quantum_action}). Unfortunately, it is not possible to
determine analytically the exact non-trivial time-dependent
solutions (instantons) of the above non-linear partial differential
equation. The physical process of tunneling may however be
understood as a two-stage process~[\onlinecite{LL}]. The first stage (small times)
corresponds to the local tunneling of the CDW at the impurity
position by one period, $a$. This stage is described by the
backscattering term which is local in space. We now assume that
the non-trivial dynamics of the phase at the impurity are essential
only at large times (large and small times will be defined below).
As a result the jump of the phase, at the impurity, during the first
stage is described by the Ansatz: $\varphi(\tau) = 2 \pi
\theta(\tau)$, where $\theta(\tau)$ is the Heaviside function. The
action describing the first stage is then linear in time and in
the impurity strength:
\begin{eqnarray}
\label{S1}
s_1 = W_b \beta.
\end{eqnarray}
The second stage (large times), corresponds to the adjustment of
the crystal, at large distances from the impurity, to the jump of
the phase at the impurity. It is described by assuming that the
local effect of the impurity is irrelevant, {\it i.e.} that one
can set $W_b = 0$. Then, the corresponding effective action may be
derived from Eq.~(\ref{quantum_action}) (with $W_b = 0$) by tracing out
all remaining gapless modes away from the impurity position. This
leads to an effective action for the phase at the impurity:
\begin{eqnarray}
\label{S2}
s_2  = { 1 \over 2 \beta} \sum_\omega |\varphi_\omega|^2 J_{\omega},
\end{eqnarray}
where the spectral function $J_\omega$ contains the effect of the infinite number of gapless
modes which have been traced out. Without the Coulomb interaction
this spectral function reads:
\begin{eqnarray}
\label{Kernel}
J_{\omega}^{-1} =
\int {d{\bf q} \over (2\pi)^2} {1 \over C\omega^2 + Y {\tilde q}^2}.
\end{eqnarray}
where ${\tilde q}^2 = q_x^2 + \alpha q_y^2$.

\subsubsection{The case of large impurity concentration}

In this case wires are elastically weakly coupled, $\al \ra 0$. This case
is worth examining first because it relates our approach to known results in the literature on 1D disordered interacting systems. Eq.~(\ref{Kernel}) leads to $J_{\omega} = 4 \pi b \sqrt{YC} |\omega|$ and therefore to an action of Eq.~(\ref{S2}) which is non-local in time:
\begin{eqnarray}
s_2 = \int {d\tau d\tau'} \left( {\varphi(\tau) - \varphi(\tau') \over \tau - \tau'} \right)^2,
\nonum
\end{eqnarray}
and is similar to the one of Caldeira and Leggett in the frame of quantum dissipation.
This terminology might be misleading here, as the dissipation arising in the
effective action is not related to an external bath coupled to the
system (phonons). It is intrinsically related to the incommensurate CDW, {\it i.e.}
the action is that of generic gapless modes. It may be understood by thinking about
the CDW away from the impurity as an effective (internal) bath
governing the dynamics of the phase at the impurity. Tracing out
the internal degrees of freedom away from the impurity, while
keeping the phase at the impurity fixed, leads to non-trivial
dynamics for the latter. This procedure is well-known in 1D disordered systems, {\it cf.}
Refs.~[\onlinecite{KF,FG,BM2}], the first extension to quasi-1D systems appearing in Ref.~[\onlinecite{BraNat}].
Because the phase is bounded along the time trajectory, {\it i.e.} $\Delta \varphi = 2\pi$, this part of the action is logarithmic in time:
\begin{equation}
s_2 = b~\sqrt{YC}~ \ln(l / u \beta),
\label{S2_1D}
\end{equation}
where $u=\sqrt{Y/C}$ is the velocity of the collective electron structure and $l / u$ an upper-time cut-off preventing the divergency of the action.

With the help of Eqs.~(\ref{S1}) and (\ref{S2_1D}), minimizing the total action $s=s_1+s_2$ with respect to $\beta$, the optimal time reads:
\begin{equation}
\label{tunn_time_1D}
\beta_{\mathrm{opt}} = b~\sqrt{YC} / W_b,
\end{equation}
and the optimal action reads:
\begin{equation}
\label{Seff_1D_part}
s_{\mathrm{opt}} = b~\sqrt{YC}~\ln(l W_b / b Y).
\end{equation}
In Eq.~(\ref{Seff_1D_part}), we have assumed that the logarithm gives
the major contribution with respect to unity. This is effectively
the case, as the inequality: $W_b l /b  \gg Y$, is equivalent to
the requirement that we are in the strong pinning regime. The
latter also implies that the impurity strength and bulk modulus
are determined by local electrostatics: $W_b \sim e^2/\kappa b$
and $Y \sim e^2/\kappa b$, as may be check by dimensional
arguments. The arguments of the logarithm therefore corresponds
to: $W_b l / b Y = l/b \gg 1$.

The second check deals with our initial assumption that at small
times, the non-trivial dynamics of the phase at the impurity
position are irrelevant. We may check that this is the case for an
infinite, delta-function-like, back-scatterer ($W_b \ra \infty$).
In this case, the tunneling is instantaneous and the phase
effectively jumps by $2 \pi$, {\it i.e.} by one period at the
impurity position. The Ansatz: $\varphi(\tau) = 2 \pi
\theta(\tau)$, where $\theta(\tau)$ is the Heaviside function, is
then perfectly justified. This is also the case if the optimal time is
smaller or equal to the smallest time-scale of the problem:
$\beta_{\mathrm{opt}} \leq a/u$. Using again the fact that the
plasmon velocity is related to our parameters with the help of: $u
= \sqrt{Y / C}$, this condition reads: $W_b \geq  Y b /a$. The lower
boundary is satisfied in the case of strong individual pinning,
$W_b \sim Y \sim e^2/\kappa b$, for sufficiently dilute electronic systems (hence
large $r_s$, see below).

Finally, we notice that the phase-phase correlation function in
the absence of disorder scales as: $<\varphi^2> = \hbar / b \sqrt{Y
C}$ (where $\hbar$ has been restored for clarity). The weakness of
quantum fluctuations, on which our starting semi-classical approach
was based, implies that: $b \sqrt{Y C}
\gg \hbar$. This condition fulfills the requirement that, in
Eq.~(\ref{Seff_1D_part}), the optimal action, $s_{\mathrm{opt}} \gg
1$. Because $u = \sqrt{Y/C}$, and using again the fact that: $Y
\sim e^2/\kappa b$, we see also that: $b \sqrt{Y C} / \hbar =
e^2/\kappa / \hbar u  = U_H / \epsilon_F $, where $U_H  =
e^2/\kappa a$ is the Coulomb energy scale and $\epsilon_F = \hbar
u /a$ is the kinetic energy scale ($a$ is the average distance
between electrons along a wire). We may therefore introduce the
well-known parameter $r_s$ which is defined as:
\be
\label{rs}
r_s \equiv {U_H \ov \eps_F} = b \sqrt{Y C} / \hbar \gg 1.
\ee
It follows from Eq.~(\ref{rs}), that the optimal action of Eq.~(\ref{Seff_1D_part}),
may be re-expressed as:
\begin{eqnarray}
\label{Seff_1D}
s_{\mathrm{opt}} = r_s \ln ({ l / b}),
\end{eqnarray}
where we have used the fact that $W_b \sim Y$ and returned $\hbar$ to unity.
It is crucial to notice that this action is non-WKB like because
the strength of the barrier, $W_b$, is in the logarithm, as known
from Larkin and Lee in the strictly 1D case~[\onlinecite{LL}]. The difference with the 1D case here is that the inter-chain distance, $b$, appears in the
logarithm instead of the average electron distance, $a$, in 1D.
From Eq.~(\ref{xi_x_def1}) and (\ref{Seff_1D}), this yields the localization length in the large impurity
concentration case:
\be
\xi_x = {1 \ov r_s b N \ln(1 / N b^2)}, \quad N \gg N_s,
\label{xi_1D_L}
\ee
where $N=1/lb$ has been used.

We include now the long-range Coulomb. Following Eq.~(\ref{eps_2D}) and the discussion below it, the inclusion of the Coulomb interaction amounts to replace: ${\tilde q}^2 = q_x^2 + \al q_y^2 \ra q_x^2$ by ${\tilde q}^2 \veps_2({\bf q})= {\tilde q}^2 + q_x^2 / r_D q \ra q_x^2 [ 1 + 1/ r_D q]$, where $q^2 = q_x^2 + q_y^2$.
Because $q r_D \ll 1$ and at large distances along the chains, $q_x \ll q_y$, this reduces to the non-analytic shift: ${\tilde q}^2 \ra q_x^2 / r_D |q_y|$ in Eq.~(\ref{Kernel}), which reads:
\begin{eqnarray}
\nonum
J_{\omega}^{-1} =
\int {d{\bf q} \over (2\pi)^2} {1 \over C\omega^2 + Y q_x^2 / r_D |q_y|}.
\end{eqnarray}
It is straightforward to show that the integration yields: $J_\om = 3 \pi b \sqrt{YC} |\om|$, which is the above result up to a numerical factor. In 2D and in the large impurity concentration case, the Coulomb interaction does not modify the result of Eq.~(\ref{xi_1D_L}).

The case where $N \ll N_s$ requires stronger inter-wire couplings and will be considered next.

\subsubsection{The case of small impurity concentration}

In the small-impurity concentration case the inter-wire coupling is crucial, {\it i.e.} it gives rise to the non-trivial scale $l_s$. For coupled wires, we are only aware of the results of Ref.~[\onlinecite{BraNat}] dealing with the equivalent 3D geometry. In our 2D case, including inter-chain interactions, and focusing on large-distances along the chains, $q_x \ll \sqrt{\al} q_y$, Eq.~(\ref{Kernel}) reads:
\begin{eqnarray}
\nonum
J_{\omega}^{-1} =
\int {d{\bf q} \over (2\pi)^2} {1 \over C \omega^2 + \al Y q_y^2}.
\end{eqnarray}
This yields:
\begin{eqnarray}
\label{Kernel_2D}
J_{\omega} =   {E_s  \over \ln (E_s / b \sqrt{Y C} |\omega|)},
\end{eqnarray}
where $E_s \propto e^2/\kappa l_s$ is the soliton energy.
The total action, $s=s_1+s_2$, in the logarithmic approximation, therefore reads:
\begin{eqnarray}
s = {W_b}\beta + {E_s \beta \over \ln (E_s \beta / b \sqrt{Y C})}.
\nonum
\end{eqnarray}
The logarithmic factor provides a minimum to the action which,
in the strong pinning regime, $W_b \gg E_s$, is then given by:
\begin{eqnarray}
\label{Seff_2D}
s_{\mathrm{opt}} = b \sqrt{ YC}~{W_b \over E_s} = r_s~{W_b \over E_s},
\end{eqnarray}
where Eq.~(\ref{rs}) has been used.

Next, we include the effect of the Coulomb interaction shifting ${\tilde q}^2$ to ${\tilde q}^2 \epsilon_2({\bf q})
= {\tilde q}^2 + q_x^2 / r_D q$, {\it cf.} Eq.~(\ref{eps_2D}) and the discussion following it.
At large distances along the chains, $q_x \ll \sqrt{\alpha} q_y$, the shift reduces to ${\tilde q}^2 \ra \al q_y^2 + q_x^2 / r_D |q_y|$. The kernel of Eq.~(\ref{Kernel}) then reads:
\begin{eqnarray}
\nonum
J_{\omega}^{-1} =
\int {d{\bf q} \over (2\pi)^2} {1 \over C \omega^2 + \al Y q_y^2 + Y q_x^2 / r_D |q_y|}.
\end{eqnarray}
The integrations yield:
\begin{eqnarray}
\label{Kernel_2D_U}
J_{\omega} =  {E_s \over 1 - (b \sqrt{Y C} |\omega| / E_s)^{1/2}}.
\end{eqnarray}
Eq.~(\ref{Kernel_2D_U}) shows that, in the presence of the Coulomb interaction, the dynamics enter the kernel in a perturbative way with respect to the static part $E_s$. This was not the case in 1D as well as in 2D without the long-range Coulomb interaction, {\it cf.} the logarithmic factor in Eq.~(\ref{Kernel_2D}). The expansion of the kernel, up to second order, leads to the following total action:
\begin{eqnarray}
s = {W_b} \beta + [b E_s \sqrt{YC} \beta ]^{1/2} + b \sqrt{YC}~\ln (l_s /u\beta),
\nonum
\end{eqnarray}
where our cut-off is now $l_s$. The optimal time is found to be:
\begin{equation}
\label{tunn_time_2D_2}
\beta_{\mathrm{opt}} = {b \sqrt{YC} \over W_b}.
\end{equation}
which leads to the following optimal action:
\begin{eqnarray}
s_{\mathrm{opt}} = b\sqrt{ YC} + b\sqrt{ YC} \left( {E_s  \over W_b} \right)^{1/2} +
b\sqrt{YC}~\ln({l_s W_b \over b Y}).~~~
\nonum
\end{eqnarray}
In the strong pinning regime, $W_b \sim Y$, and with $l_s \gg b$, the action is dominated by the logarithmic contribution which, up to a numerical factor, reads:
\begin{eqnarray}
\label{Seff_2D_U}
s_{\mathrm{opt}} =  r_s~\ln({l_s / b}).
\end{eqnarray}
where Eq.~(\ref{rs}) has been used as well as the strong pinning result: $W_b \sim Y$. The long-range Coulomb therefore returns us to the logarithmic action which is non WKB-like.

From Eqs.~(\ref{xi_x_def2}) and (\ref{Seff_2D_U}), the localization length in the small impurity concentration case reads:
\be
\xi_x = {1 \ov r_s b N_s \ln(1 / N_s b^2)}, \quad N \ll N_s,
\label{xi_1D_S}
\ee
where $N=1/lb$ has been used ($N_s = 1 / l_s b$).

\subsubsection{General expression and remark}

Eqs.~(\ref{xi_1D_L}) and (\ref{xi_1D_S}) yield the general expression for the localization length:
\be
\xi_x^{-1} = r_s~b~{\mathrm{max}}\{N,N_s\}~\ln({1  \ov {\mathrm{max}}\{N,N_s\} b^2}),
\label{xi_expression}
\ee
which we will use in the following section.

As a final note, we follow Ref.~[\onlinecite{BraNat}], where it has been mentioned that the collective dynamics of the CDW may include a contribution from the amplitude, $|\Delta(\tau)|$. This contribution arises because the order parameter is complex: $\Delta(\textbf{r},t) = |\Delta| \exp(i\varphi)$. At zero temperature the amplitude mode is frozen but in VRH we may be interested in reaching the thermally activated regime. In this case, there is an additional regular kinetic energy contribution $\propto I \omega^2$, where the momentum of inertia $I$ depends on fluctuations of the amplitude. This brings an additional contribution to the total action of Eq.~(\ref{Seff_2D_U}): $\propto I/\beta$. Eq.~(\ref{Seff_2D_U}) would then be valid provided that: $I \ll b^2 YC  / W_b$. In the other case, $I \gg b^2 YC / W_b$, the final action reads:
\begin{eqnarray}
\label{Seff_2D_I}
s_{\mathrm{opt}} = 2 \sqrt{I W_b} + b\sqrt{ YC}~\ln({L \sqrt{W_b} \over u \sqrt{I}}),
\end{eqnarray}
where the second term is a correction and $L={\mathrm{min}}\{l,l_s\}$. Notice that the first term, with a square-root dependence on the barrier, is WKB-like. In the following, we will assume that the amplitude mode has a negligeable contribution and we will use the final expression for the localization length, Eq.~(\ref{xi_expression}), based on the non WKB-like actions. Depending on the system under consideration the WKB contribution may however play a significant role. One would then have to use Eq.~(\ref{Seff_2D_I}) together with Eqs.~(\ref{xi_x_def1}) and (\ref{xi_x_def2}) to derive the localization length.


\section{Linear VRH laws}
\label{linearVRH}

This section deals with the hopping laws of the quasi-1D
Mott-Anderson insulators in the linear response regime. Our results depend crucially on the Coulomb gap shapes derived in Sec.~\ref{coulombgap} and on the impurity-dependence of the longitudinal localization length derived in Sec.~\ref{tunneling}. Notice that all transport laws below, and especially the parameter characteristic of these laws, are given up to a numerical coefficient of the order of unity.

Recall that the semi-phenomenological arguments initially introduced by Mott in order
to derive the dc conductivity, $\sigma(T)$, are based on minimizing (with respect to
the coordinates $x$ and $y$) the toy-action:
\begin{equation}
\label{toy_s}
S = 2 x/ \xi_x + 2 y / \xi_y + \Gamma / T,
\end{equation}
where the quantum and classical parts are related by a Fermi's
Golden rule involving the density of localized states:
\begin{equation}
\label{GoldenRule} \nu \Gamma \mathcal{S} \approx 1,
\end{equation}
where $\mathcal{S} = x y$ is the area. Eqs.~(\ref{toy_s}) and~(\ref{GoldenRule})
describe the process with the help of which an electron may hop from a site $i$ to the site $j$
within a disorder band with a constant density of localized states,
$\nu$. The first two terms in Eq.~(\ref{toy_s}) represent the
overlap between the two states separated by $x$ and $y$.
Because the system is disordered the energies of these states are
different and a phonon has to be involved in the process of
hopping. The thermal energy, $\Gamma$, is then
determined with the help of Eq.~(\ref{GoldenRule}) and corresponds to
the average energy spacing between the localized states involved
in the hopping: $\Gamma = 1/\nu \mathcal{S} = 1/\nu x y$. Introducing this value of $\Gamma$ in Eq.~(\ref{toy_s}) and minimizing the total action with respect to $x$ and $y$ yields the optimal hopping distances:
\be
x_{\mathrm{opt}} = \xi_x~(T_{M}/T)^{1/3}, \quad y_{\mathrm{opt}} = \xi_y~(T_{M}/T)^{1/3},
\nonum
\ee
where the parameter $T_M$ reads:
\be
T_M = 1 / \nu \xi_x \xi_y.
\label{TM_general}
\ee
This yields an optimal action: $S_{\mathrm{opt}} \propto (T_{M}/T)^{1/3}$.
The corresponding d.c. conductivity, $\sigma \propto \exp(-S_{\mathrm{opt}})$, is the Mott law for variable-range hopping and reads:
\begin{equation}
\sigma_M(T) = \sigma_0(T) \exp[-(T_{M}/T)^{1/3}],
\label{DCMott}
\end{equation}
where $\sigma_0$ is a temperature-dependent pre-factor (the temperature-dependence
arising from the phonon-scattering time which is a power-law).
This law may be straightforwardly extended to any dimension. It may have a disorder-dependence through the constant DOS of the disorder band, $\nu$:
\begin{subequations}
\label{nu}
\bea
&&\nu \propto {N \ov e^2 / \kappa l} = {\kappa \ov e^2 b}, \qquad ~~~~ N \gg N_s,
\label{nu_L} \\
&&\nu \propto {N \ov e^2 / \kappa l_s} = {\kappa \ov e^2 b}~{N \ov N_s}, \quad N \ll N_s.
\label{nu_S}
\eea
\end{subequations}
This estimate was derived on the basis that we have $N$ localized states per unit volume, each with an energy $e^2 / \kappa l$ in the limit $N \gg N_s$ (charge excitations extend over segments between impurities, $l$) and $e^2 / \kappa l_s$ in the limit $N \ll N_s$ (charge excitations have their length-scale, $l_s$). This shows that $\nu$ grows with $N$ in the small impurity concentration case and saturates when $N$ becomes larger than $N_s$. An additional dependence may come from $\xi_x$. In particular from Eqs.~(\ref{TM_general}) and (\ref{xi_expression}), the parameter of the Mott-law for 2D quasi-1D systems reads:
\begin{subequations}
\label{TM_expression}
\bea
&&T_M = {r_s b N \ln(1 / N b^2) \ov \nu \xi_y}, \quad ~~ N \gg N_s,
\label{TM_L} \\
&&T_M = {r_s b N_s \ln(1 / N_s b^2) \ov \nu \xi_y}, \quad N \ll N_s,
\label{TM_S}
\eea
\end{subequations}
where $N=1/lb$ is the impurity-density, $b$ the inter-wire distance and $N_s$ has been defined by Eqs.~(\ref{N_S}) and (\ref{N_L}). It is interesting to notice, from Eq.~(\ref{TM_L}), that in the large impurity concentration case there is a linear dependence on $N$ arising from $\xi_x$ ($\nu$ is constant from Eq.~(\ref{nu_L})).
On the other hand, in the small impurity concentration case, Eq.~(\ref{TM_S}), $\xi_x$ saturates and we have $T_M \sim 1/N$ from the $N-$dependence of $\nu$, Eq.~(\ref{nu_S}). In this case, the conductivity increases with increasing disorder.

When the long-range Coulomb interaction is taken into account, the
latter leads to a depletion of low-energy states in $\nu$. As we
have seen in Sec.~\ref{coulombgap}, this depletion corresponds to a
soft Coulomb gap $g(\veps)$ in $\nu$. Following Efros and
Shklovskii, in the presence of this Coulomb gap, one has to
replace Eq.~(\ref{GoldenRule}) by:
\begin{equation}
\label{GoldenRule_CG}
g(\Gamma) \Gamma \mathcal{S} \approx 1.
\end{equation}
With the help of Eq.~(\ref{gES}), corresponding to a usual ES Coulomb gap shape, and the arguments above, the d.c. conductivity is the Efros-Shklovskii law for VRH:
\begin{equation}
\sigma_{ES}(T) = \sigma_0(T) \exp[-(T_{ES}/T)^{1/2}],
\label{DCES}
\end{equation}
where the parameter $T_{ES}$ depends on the specific Coulomb
interaction among charge carriers:
\be
T_{ES} = {e^2 \ov \kappa \sqrt{\xi_x \xi_y}}.
\label{TES_general}
\ee
Substituting the expression of the localization length, Eq.~(\ref{xi_expression}),
this parameter reads:
\begin{subequations}
\label{TES}
\bea
&&T_{ES} = {e^2 \ov \kappa} \sqrt{r_s N b \ln(1 / N b^2) \ov \xi_y}, \quad ~~ N \gg N_s,
\label{TES_L} \\
&&T_{ES} = {e^2 \ov \kappa} \sqrt{r_s N_s b \ln(1 / N_s b^2) \ov \xi_y}, \quad N \ll N_s.
\label{TES_S}
\eea
\end{subequations}
In the large impurity concentration case, $N \gg N_s$, this parameter increases with the impurity concentration, $N$, which leads to a decreasing conductivity as a function of disorder. This impurity-dependence originates from $\xi_x$. In the small impurity concentration case, $N \ll N_s$, $\xi_x$ saturates and $T_{ES}$, as well as the corresponding conductivity, become disorder-independent.

At higher energies, the usual Coulomb gap of Eq.~(\ref{gES}) crosses over to the unusual Coulomb gap of Eq.~(\ref{g1}). The latter yields a conductivity that we denote as: $\sigma_1$, and reads:
\begin{equation}
\sigma_1(T) = \sigma_0(T) \exp[-(T_1/T)^{3/5}].
\label{DC1}
\end{equation}
The unusual exponent, {\it i.e.} $3/5$ instead of $1/2$, is related to the unusual Coulomb gap of Eq.~(\ref{g1}).
Furthermore, the parameter $T_1$ reads:
\be
T_1 = {e^2 \over \left(b \kappa^2 \kappa_x \xi_x \xi_y \right)^{1/3}}.
\label{T1_general}
\ee
Substituting the expression of $\xi_x$, Eq.~(\ref{xi_expression}), and of $\kappa_x$, Eq.~(\ref{kappax}), yields:
\begin{subequations}
\label{T1}
\bea
&&T_1 = {e^2  \ov \kappa b}~N b^2~\left[ {r_s b \ln(1/Nb^2) \ov \xi_y} \right]^{1/3}, ~~~ N \gg N_s,~~~~
\label{T1_L} \\
&&T_1 = {e^2 b \ov \kappa} \left[ {r_s b N^2 N_s \ln(1/N_s b^2) \ov \xi_y} \right]^{1/3}, ~ N \ll N_s.~~~~
\label{T1_S}
\eea
\end{subequations}
The $T_1$ parameter depends on disorder through both $\xi_x$ and $\kappa_x$, in the large-impurity concentration regime. On the other hand, $\xi_x$ saturates in the small impurity concentration case and the disorder dependence of $T_1$ originates then only from $\kappa_x$. In both cases this disorder-dependence of $T_1$ yields a conductivity, $\sigma_1$, which decreases with increasing disorder.

We then focus on the peculiar $2/3$ Coulomb-gap sector in the small impurity case, {\it cf.} Eq.~(\ref{g2}).
This Coulomb-gap yields:
\begin{equation}
\sigma_2(T) = \sigma_0(T) \exp[-(T_2/T)^{5/11}], \quad N \ll N_s.
\label{DC2}
\end{equation}
The unusual exponent, {\it i.e.} $5/11$ instead of $1/2$, is again related to the unusual Coulomb gap (here $\propto \veps^{2/3}$) of Eq.~(\ref{g2}). Furthermore, the parameter $T_2$ reads:
\be
T_{2} = e^2 \left[ {\alpha r_D \over \xi_x^3 ~\xi_y^3} \right]^{1/5}, \quad N \ll N_s.
\label{T2_general}
\ee
Substituting the expression for $\xi_x$, Eq.~(\ref{xi_expression}), this parameter reads:
\be
T_2 = e^2 \left[ {\alpha r_s^3 b^4 N_s^3 \ln^3(1/N_sb^2) \over \xi_y^3} \right]^{1/5}, \quad N \ll N_s.
\ee
This parameter, as well as the corresponding conductivity, $\sigma_2$, are disorder-independent.

\begin{figure}
\includegraphics[width=6cm,height=3cm]{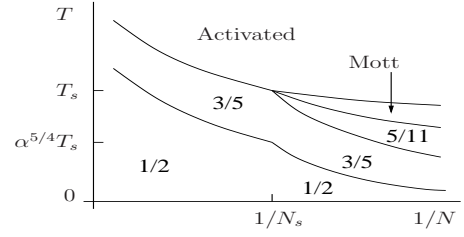}
\caption{ \label{PD_2D} The generic phase diagram of a 2D quasi-1D system as a
function of temperature $T$ and the inverse average impurity concentration $1/N$ [$T_s \propto E_s$ the
soliton energy]. This phase diagram displays the VRH laws with exponents
$1/2$, $3/5$ and $5/11$ as well as Mott law and activation at higher temperatures.}
\end{figure}

Finally, at high energies (within the constant DOS $\nu$), these VRH laws cross-over to the so called nearest-neighbor hopping (NNH) or activated law, see Ref.~\onlinecite{ES_book} for a review. The latter reads:
\begin{equation}
\label{DCNNH}
\sigma_{NNH}(T) = \sigma_0(T) \exp[-E_{NNH}/T],
\end{equation}
where the NNH energy reads:
\begin{subequations}
\bea
\label{NNH}
&&E_{NNH} =  e^2 b N / \kappa, \quad ~ N \gg N_s,
\label{NNH_L} \\
&&E_{NNH} = e^2 b N_s / \kappa, \quad N \ll N_s.
\label{NNH_S}
\eea
\end{subequations}
These results are summarized on the temperature $-$ impurity-concentration phase-diagram of Fig.~\ref{PD_2D}. The crossover lines between the different laws in Fig.~\ref{PD_2D} are determined with the help of the following arguments.

The expression of $\veps_1$, Eq.~(\ref{eps1}), determines the crossover temperature lines $T_1(N)$ between
the ES and the $3/5-$laws [$\veps_1 = T_{c1} (T_{ES}/T_{c1})^{1/2}$] with $T_{ES}$ given by
Eq.~(\ref{TES}). This crossover line reads
\begin{subequations}
\bea
\label{Tc1}
&&T_{c1} = {e^2 \over \kappa b}~(N b^2)^{7/2}~\sqrt{{\xi_y \over r_s b \log({1 \ov N b^2})}},
~~~ N \gg N_s,~~~~
\label{Tc1_L} \\
&&T_{c1} = {e^2 \over \kappa b^2}~(N b^2)^4~\sqrt{{\xi_y \over r_s b N_s \log({1 \ov N_s b^2})}},
~ N \ll N_s.~~~~
\label{Tc1_S}
\eea
\end{subequations}
In the case of large impurity concentration, the expression of $\Delta_1$, Eq.~(\ref{Delta_1}) determines the total width of the Coulomb gap and depends on the constant DOS $\nu$. This width determines the high
temperature crossover line $T_{c2}(N)$, [$\Delta_1 = T_{c2} (T_{1}/T_{c2})^{3/5}$],
between the $3/5$-law and nearest-neighbor-hopping, {\it cf.} Eq.~(\ref{NNH}), for $N \gg N_s$. With the help of Eq.~(\ref{DC1}), $T_{c2}(N)$ reads:
\bea
T_{c2}(N) =&& {e^2 \over \kappa b}~N b^2\left[ {e^2 \nu b \ov \kappa} \right]^{5/4}~
\sqrt{{\xi_y \ov r_s b \ln(1/Nb^2)}},
\nonum \\
&& N \gg N_s.
\label{Tc2_L}
\eea
Using the estimation for $\nu$, Eq.~(\ref{nu_L}), Eq.~(\ref{Tc2_L}) may be re-expressed as:
\be
T_{c2}(N) = {e^2 \over \kappa b}~N b^2~\sqrt{{\xi_y \ov r_s b \ln(1/Nb^2)}},~
N \gg N_s.
\label{Tc2_L2}
\ee
From Eqs.~(\ref{Tc1}) and~(\ref{Tc2_L}), we see that: $T_{c2}(N)/T_{c1}(N) = 1/(b^2 N)^{5/2} \gg 1$, so that at the boundary $N \sim N_s$, the corresponding crossover temperatures are separated by the large dimensionless factor $(l_s/b)^{5/2} = 1/\al^{5/4}$. Moreover, we define $T_s \equiv T_{c2}(N_s)$ which reads, up to a logarithmic factor:
\begin{equation}
\label{T_s}
T_s \equiv T_{c2}(N_s) \propto E_s,
\end{equation}
where: $E_s = e^2 / \kappa l_s$, is the creation energy of a soliton. With these notations we have:
\begin{equation}
\label{T_c1_ls}
T_{c1}(N_s) \propto \alpha^{5/4}T_s \ll T_s.
\end{equation}
We could also show that in this large impurity concentration case there is no room for the Mott law of Eq.~(\ref{TM_L}), {\it i.e.} the crossover line from the $3/5-$law to the Mott law coincides with the crossover line from the Mott law to the activated- or NNH- law.

In the small impurity concentration case we have already given the expression for the $T_1-$line in Eq.~(\ref{Tc1_S}), which matches smoothly Eq.~(\ref{Tc1_L}) at $N \sim N_s$. On the other hand, the $T_2-$ line splits into three crossover lines thereby opening two sectors for the new $5/11-$law and the Mott law. The energy $\veps_2$, Eq.~(\ref{eps2}), determines the temperature crossover-line: $T_{c2}^a(N)$, [$\veps_2 = T_{c2} (T_{1}/T_{c2})^{3/5}$],
between the $3/5-$law and the $5/11-$laws for $N \ll N_s$. With the help of Eq.~(\ref{T1_S}) for the parameter $T_1$ and Eq.~(\ref{eps2}), $T_{c2}^a(N)$ reads:
\begin{equation}
\label{Tc2_aS}
T_{c2}^a(N) = {e^2 \over \kappa l_s} \left[ {N \over N_s} \right]^{11/4}
\sqrt{ { \xi_y \over r_s b \log(1 / N_s b^2) } }, \quad N \ll N_s.
\end{equation}
which matches smoothly Eq.~(\ref{Tc2_L2}) at $N \sim N_s$ and decreases more abruptly with $1/N$ for $N \ll N_s$.

At the next crossover-line: $T_{c2}^b(N)$, the $5/11-$law crosses over to the Mott VRH law
with the parameter Eq.~(\ref{TM_S}). This crossover-line reads:
\be
\label{Tc2_bS}
T_{c2}^b(N) = {e^2 \over \kappa l_s}~\left[ {N \over N_s} \right]^{5/2}~\sqrt{ \xi_y \over r_s b \log(1 / N_s b^2)},~~ N \ll N_s,
\ee
which matches smoothly Eqs.~(\ref{Tc2_L2}) and (\ref{Tc2_aS}) at $N \sim N_s$.
Finally, at higher temperatures a crossover-line bridges the Mott law of Eq.~(\ref{TM_S}) with the NNH law of Eq.~(\ref{NNH_S}):
\be
\label{Tc2_cS}
T_{c2}^c(N) = {e^2 \over \kappa l_s}~\left[ {N \over N_s} \right]^{1/2}~\sqrt{ \xi_y \over r_s b \log(1 / N_s b^2)}, \quad N \ll N_s,
\ee
which matches smoothly Eqs.~(\ref{Tc2_L2}), (\ref{Tc2_aS}) and (\ref{Tc2_bS}) at $N \sim N_s$.


\section{Non-linear VRH laws}
\label{nonlinearVRH}

This section deals with the hopping laws of the quasi-1D
Mott-Anderson insulators in the non-linear response regime. Our results depend crucially on the Coulomb gap shapes derived in Sec.~\ref{coulombgap} and on the impurity-dependence of the longitudinal localization length derived in Sec.~\ref{tunneling}. Our arguments follow closely those of the linear-response regime of Sec.~\ref{linearVRH}. Notice that all transport laws below, and especially the parameter characteristic of these laws, are given up to a numerical coefficient of the order of unity. Notice also that our present arguments do not allow us to determine the pre-factor of the current; we therefore focus only on its main exponential dependence.

The VRH laws derived in Sec.~\ref{linearVRH} are valid in the linear response (or ohmic) regime where the current is linear in the applied electric field, $j=\sigma(T) \mathcal{E}$. At a given temperature, upon increasing the electric
field, a crossover should take place, below the threshold for the sliding of the electronic crystal, to a non-linear regime. In the frame of doped semiconductors, such a transition has been studied in Ref.~\onlinecite{Boris} by extending the Mott argument presented in the previous subsection on the linear VRH laws. Such an extension
amounts to replace, in Fermi's Golden rule of Eq.~(\ref{GoldenRule}), the hopping energy $\Gamma$ by the energy provided by the electric field, $e \mathcal{E} r$, during the motion of an electron along a distance $r$
from the initial localized state. In d dimensions, the optimal hopping distance is therefore given by: $r_{\mathrm{opt}} = 1/(\nu e \mathcal{E})^{1/1+d}$. Substituting this value in the tunneling probability $\propto \exp(-2r/\xi)$
yields:
\begin{equation}
\label{current_MS}
j_{MS}(\mathcal{E}) \sim \exp[-({\mathcal{E}_{MS} \ov \mathcal{E}})^{1 \over 1+d}],~~
\mathcal{E}_{MS} = {1 \over \nu e \xi^{1+d}}.
\end{equation}
where the index $MS$ refers to Mott-Shklovskii, see also Ref.~\onlinecite{Nattermann}. The crossover between the linear and non-linear regimes takes place when: $(\mathcal{E}_{MS}/\mathcal{E})^{1/1+d} = (T_M/T)^{1/1+d}$, where the parameter of the Mott law: $T_M = 1/\nu \xi^d$, in d-dimensions. At a given temperature, the threshold field for the non-ohmic regime is given by: $\mathcal{E}_c = (\mathcal{E}_M /T_M) T $, that is:
\begin{equation}
\label{E_nonohmic}
\mathcal{E}_{c} = {T \over e \xi},
\end{equation}
in all dimensions. This returns us to the arguments of the Introduction, see Eq.~(\ref{constraint_non_linear}) and discussion around it.

It is straightforward to generalize such results in the presence of a Coulomb gap. We need simply to replace the constant DOS, $\nu$, in Fermi's Golden rule by the corresponding Coulomb gap: $g(\varepsilon)$, evaluated at the energy of the charge
excitation: $\varepsilon = eEr$. The generic form of the Coulomb-gap (CG) in isotropic system is given by Eq.~(\ref{CG_iso}). This leads to an optimal hopping distance independent of dimensionality: $r_{\mathrm{opt}} = (e^2 / \kappa e \mathcal{E})^{1/2}$. The corresponding current is therefore given by:
\begin{equation}
\label{current_0}
j_0(\mathcal{E}) \sim \exp[-(\mathcal{E}_0/\mathcal{E})^{1 \over 2}],~~
\mathcal{E}_0 = {e^2 \over \kappa e \xi^2},
\end{equation}
Strictly speaking, Eq.~(\ref{current_0}) is valid for $d > 1$. In $d=1$, there is a logarithmic Coulomb gap: $g(\varepsilon) = \nu/\log(e^2 N b/\kappa |\varepsilon|)$, see Ref.~\onlinecite{FTS} for a similar derivation of this law in the ohmic regime. For the 1D case, the parameter of Eq.~(\ref{current_0})
is therefore given by:
\begin{equation}
\label{current_0_1D}
\mathcal{E}_0 = { \log[(e^2 N b/ \kappa)\sqrt{\nu/ \mathcal{E}}] \over \nu e \xi^2}, \quad d=1,
\end{equation}
in the logarithmic approximation where the energy of the charge excitation has been taken equal to
$\varepsilon = e \mathcal{E} r_{\mathrm{opt}}$ and $r_{\mathrm{opt}} = 1/\sqrt{\nu e \mathcal{E}}$ is the optimal hopping distance in 1D.

These results may be extended to quasi-1D systems. For simplicity, we assume that the {\it electric field is parallel
to the chains}. More general results with the two components of the field can be derived in the same way. We also focus on the 2D case. Then,
Fermi's Golden rule of Eq.~(\ref{GoldenRule}) is generalized to: $\nu (e \mathcal{E}_x x) x y \sim 1$. The tunneling action is anisotropic and takes the usual form $S=2x/\xi_x + 2y/\xi_y$. Substituting
the $x-$component of the hopping length in this action and minimizing the resulting expression with respect to
$y$ yields to the optimal hopping distances. Substituting the latter in the action we finally obtain a current:
\begin{equation}
\label{current_MS_2}
j_{MS}(\mathcal{E}) \sim \exp[-({\mathcal{E}_{MS} \ov \mathcal{E}})^{1 \over 3}],~~
\mathcal{E}_{MS} = {1 \over \nu e \xi_x^2 \xi_y}.
\end{equation}
Substituting the expression for the longitudinal localization length, Eq.~(\ref{xi_expression}), this yields:
\begin{subequations}
\nonum
\bea
&&\mathcal{E}_{MS} = {r_s^2 b^2 N^2 \ln^2(1 / N b^2) \ov e \nu \xi_y}, \quad ~~ N \gg N_s,
\nonum \\
&&\mathcal{E}_{MS} = {r_s^2 b^2 N_s^2 \ln^2(1 / N_s b^2) \ov e \nu \xi_y}, \quad N \ll N_s,
\nonum
\eea
\end{subequations}
Substituting the expressions for $\nu$, Eqs.~(\ref{nu}), yields the explicit $N-$dependence of the parameter:
\begin{subequations}
\label{EMS_expression2}
\bea
&&\mathcal{E}_{MS} = {e^2 \ov \kappa b}~(N b^2)^2~{r_s^2 \ln^2({1 \ov N b^2}) \ov e \xi_y}, \quad ~~~ N \gg N_s,
\label{EMS_L} \\
&&\mathcal{E}_{MS} = {e^2 \ov \kappa b}~{N_s \ov N}~{r_s^2 (N_s b^2)^2 \ln^2({1 \ov N_s b^2}) \ov e \xi_y}, ~ N \ll N_s.~~~
\label{EMS_S}
\eea
\end{subequations}
The next step consists, as for the isotropic system, in introducing the Coulomb interaction, which amounts to replace the constant DOS, $\nu$, by the Coulomb gaps determined in Sec.~\ref{coulombgap}. For the linear Coulomb gap of Eq.~(\ref{gES}) the non-linear law follows Eq.~(\ref{current_0}) with a modified parameter:
\begin{equation}
\label{E0}
\mathcal{E}_{0} = {e \over \kappa} \left[{1 \ov \xi_x \xi_y^{{1 \over 3}}} \right]^{{3 \over 2}}.
\end{equation}
Substituting the expression for the longitudinal localization length, Eq.~(\ref{xi_expression}), this parameter reads:
\begin{subequations}
\label{E0_expression}
\bea
&&\mathcal{E}_{0} = {e \ov \kappa}~\left[{r_s b N \ln(1 / N b^2) \ov \xi_y^{1/3}}\right]^{3/2}, ~~~ N \gg N_s,
\label{E0_L} \\
&&\mathcal{E}_{0} = {e \ov \kappa}~\left[{r_s b N_s \ln(1 / N_s b^2) \ov \xi_y^{1/3}}\right]^{3/2}, ~ N \ll N_s.~~
\label{E0_S}
\eea
\end{subequations}
On the other hand, the anomalous quadratic CG of Eq.~(\ref{g1}) leads to a $3/5-$law in the non-linear regime:
\begin{equation}
\label{current_1}
j_1(\mathcal{E}) \sim \exp[-(\mathcal{E}_1/\mathcal{E})^{3/5}],
\end{equation}
where the parameter reads:
\begin{equation}
\label{E1}
\mathcal{E}_1 = {e \over \kappa} \left[ {\kappa \over \kappa_x \xi_x^4 \xi_y b } \right]^{1/3}.
\end{equation}
Substituting the expression for the longitudinal localization length, Eq.~(\ref{xi_expression}), this parameter reads:
\begin{subequations}
\label{E1_expression}
\bea
&&\mathcal{E}_{1} = {e \ov \kappa b^2}~(N b^2)^2~\left[{r_s^4 b \ln^4({1 \ov N b^2}) \ov \xi_y}\right]^{1 \ov 3}, ~~~ N \gg N_s,~~~
\label{E1_L} \\
&&\mathcal{E}_{1} = {e \ov \kappa b^2}~(N N_s^2 b^6)^{2 \ov 3} \left[{r_s^4 b \ln^4({1 \ov N b^2}) \ov \xi_y}\right]^{1 \ov 3}, N \ll N_s.~~~~~~~
\label{E1_S}
\eea
\end{subequations}
Finally, in the small impurity concentration case, the $2/3-$Coulomb gap of Eq.~(\ref{g2}) leads to a $5/11-$law for the current:
\begin{equation}
\label{current_2}
j_2(\mathcal{E}) \sim \exp[-(\mathcal{E}_2/\mathcal{E})^{5/11}], ~~ N \ll N_s,
\end{equation}
with a parameter
\begin{equation}
\label{E2}
\mathcal{E}_2 = {e \over \kappa} \left[ {\alpha r_D \over \xi_x^8 ~ \xi_y^3 } \right]^{1/5}, ~~ N \ll N_s.
\end{equation}
Substituting the expression for the longitudinal localization length, Eq.~(\ref{xi_expression}), in the small impurity case, this parameter reads:
\be
\label{E2_expression}
\mathcal{E}_2 = {e \over \kappa l_s^2}~\left[ {r_s^8 b^3 \ln^8(1/N_s b^2) \over \xi_y^3 } \right]^{1/5}, ~~N \ll N_s.
\ee
From Eqs.~(\ref{EMS_expression2}), (\ref{E0_expression}), (\ref{E1_expression}) and (\ref{E2_expression}) for the parameters and the expressions of the corresponding currents, the non-monotonous behavior of $j$ as a function of $N$ is rather clear.

We shall not detail the field crossover lines between the various laws. This can be done exactly in the same way as for the linear-response regime case. Rather than that we would like to point out that the phase-diagram, electric-field vs. impurity concentration, may be obtained from the one of the linear law by assuming that the electric field gives rise to an "effective" temperature, $T_\mathcal{E}$. This effective temperature reads:
\be
T_\mathcal{E} ~ \equiv ~ e ~\mathcal{E}~x_{\mathrm{opt}},
\ee
where $x_{\mathrm{opt}}$ is the optimal hopping distance along a wire (recall that the electric field is assumed to be parallel to the chains). For a given non-linear (NL) law of parameter $\mathcal{E}_{NL}$ and exponent $\gamma_{NL}$, the optimal hopping distance along the wires is defined as:
\be
x_{\mathrm{opt}}(N,\mathcal{E}) = \xi_x~(\mathcal{E}_{NL} / \mathcal{E})^{\gamma_{NL}},
\ee
and depends on the applied electric field and, eventually, the impurity concentration. The phase diagram, $T_\mathcal{E}(N)$ vs. $N$, therefore has a non-trivial re-scaling via the $N-$dependence of $T_\mathcal{E}$.


\section{Conclusion}
\label{conclusion}

\begin{figure}
\includegraphics[width=7cm,height=3cm]{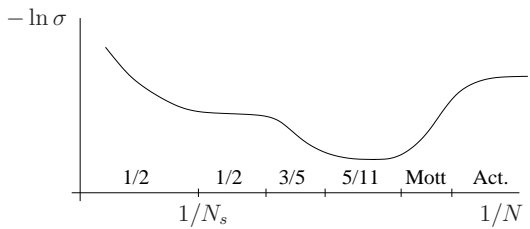}
\caption{\label{sigma_2D} The logarithm of the resistivity, in the linear response regime, versus the inverse average impurity concentration, $1/N$, for a 2D quasi-1D system at a temperature: $T \ll \alpha^{5/4} T_s$, where $T_s \propto E_s$, see Fig.~\ref{PD_2D}. The ES [Eq.~(\ref{DCES})], 3/5 [Eq.~(\ref{DC1})], 5/11 [Eq.~(\ref{DC2})], Mott [Eq.~(\ref{DCMott})] and activation [Eq.~(\ref{DCNNH})] law succeed each-other with decreasing $N$.}
\end{figure}

We attempted to construct a semi-phenomenological theory of variable-range hopping for 2D quasi-1D systems such as arrays of quantum wires in the Wigner-crystal regime. We have closely followed the phenomenological arguments of Mott, Efros and Shklovskii to derive the Coulomb gap shapes, Sec.~\ref{coulombgap}, as well as the main exponential dependence of the transport laws in the linear, Sec.~\ref{linearVRH}, and non-linear, Sec.~\ref{nonlinearVRH}, response regimes. Our approach has been supplemented with some microscopic arguments necessary to derive the impurity-dependence of the longitudinal localization length, Sec.~\ref{tunneling}. Both Coulomb gap shapes and transport laws were found to have rather unusual features with respect to known results in the field of disordered semiconductors [\onlinecite{ES_book}]. These unusual features arise because of the non-trivial dielectric properties of the systems under consideration, Sec.~\ref{dielectrics}. They are two-fold: a non-monotonous dependence of the conductivity or current as a function of disorder and a highly non-universal exponent $\gamma$. In the linear response regime, the richness of exponents is displayed on the phase diagram of Fig.~\ref{PD_2D}. Despite the fact that some exponents may be close to each-other ($1/2 = 0.50$, $3/5=0.60$, $5/11=0.45$, $1/3=0.33$), the corresponding monotonicity of the conductivity as a function of disorder enables further discrimination between the various laws. This is schematically displayed on Fig.~\ref{sigma_2D}. Moreover, 2D layers of wire arrays are experimentally accessible, to our knowledge [\onlinecite{Himpsel}], but we are unaware of systematic transport experiments in the strongly localized regime for such systems. Our theory is therefore only of predictive nature. We hope, however, that it will be of some interest to both theorists and experimentalists working in the field.

\appendix

\section{2D model of interrupted strands}
\label{appendix}

As an alternative to the use of Eqs.~(\ref{U_2D})
and~(\ref{eps_L_2D}) the ``disorder averaged'' electrostatic
potential may be derived following the results of the interrupted
strand model, see Ref.~\onlinecite{RB} of Rice and Bernasconi for the 3D case.
In this approximation, we {\it assume} that the dielectric
constant along the chains is given by Eq.~(\ref{kappax}). This
equation shows that for $L_x/r_D = l/r_D \gg 1$, the contribution
of the electronic part along the chains is much larger than the transverse part as well as the host lattice dielectric constant ($\kappa_y= \kappa \ll \kappa_x$). The problem therefore reduces to determine the electrostatic potential of a charge carrier in a layer of a quasi-one dimensional system whose
longitudinal dielectric constant is much larger than the dielectric constant of
the surrounding media. Keldysh~\cite{Keldysh} has solved a similar
problem for an isotropic layer. In the present case, the
Fourier transform of the interaction potential at $({\bf r},z)$
due to a point charge at the origin of the layer is found by usual
methods of electrostatics and reads:
\begin{equation}
\label{phiK_2D} U({\bf q},z) = {4 \pi e^2 \over \breve{q} + q}
{\cosh \left( \breve{q}b/2 + \delta \right) \over \sinh \left(
\breve{q}b + 2\delta \right)} \exp \left( q \left(b/2-z
\right)+\delta \right),
\end{equation}
where ${\bf q}=(q_x,q_y)$ is the two-dimensional reciprocal vector
of ${\bf r}$, $\breve{q}^2 = \kappa_x q_x^2 + q_y^2$ and $\delta$
is given by:
\begin{equation}
\label{gamma_2} \delta = {1\over 2}\log \left( {\breve{q}+q \over
\breve{q}-q} \right).
\end{equation}
At large distances $\breve{q}b \ll 1$, {\it i.e.} for $x \gg l$, that
will be of interest to us in the following, and for $z=0$,
Eq.~(\ref{phiK_2D}) reduces to Eq.~(\ref{U_2D}) with the following
dielectric function:
\begin{equation}
\label{kappa_keld} \veps_2({\bf q}) \approx \kappa [
1+{\kappa_x \over \kappa}{b q_x^2  \over q}].
\end{equation}
This expression is equal to Eq.~(\ref{eps_L_2D}) in the limit
$\tilde{q} = q_x \ll L_x^{-1}$, with $\kappa_x \approx \kappa
(L_x/r_D)^2$ and $r_D \approx b$. This shows that in this limit,
both approaches are equivalent.


\acknowledgements

The results presented in this manuscript were mostly derived at the William Fine TPI,
Minneapolis, during the year 2003/2004. I am indebted to S. Brazovskii and B. Shklovskii for given me then, each with his rather unique point of view on solid-state physics, important advice and comments. This manuscript has been considerably re-manipulated at the Abdus Salam ICTP, Trieste. I am sincerely grateful to V. Kravtsov and S. Scandolo for giving me the opportunity to work in such an inspiring environment.



\begin{thebibliography}{99}
\bibitem{Note:1D}
The dimension 1 is special, see Ref.~\onlinecite{FTS} for details.
\bibitem{Note:2DCDW}
In a pure 2D system, there is only quasi-long-range order. We will however still use
the terminology: CDW throughout the manuscript. As the system is strongly disordered there is no long-range order at all anyway (and it would be the same in the equivalent 3D geometry).
\bibitem{Note:phenom}
It should be noticed that going beyond the Mott or ES arguments or proving from microscopics the VRH laws are still open and challenging problems. However, progress has been made towards deriving the Coulomb gap shape, see Ref.~[\onlinecite{MI}].
\bibitem{Note:quasi-1D_parameter}
For quasi-1D systems the parameter of the VRH law reads: $T_0 = 1/\nu
\xi_x \xi_y$, where $\xi_x \propto v_F \tau_{el}$ is the
longitudinal, {\it i.e.} parallel to the chains, localization
length and $\xi_y$ is the transverse localization length.
Assuming $\xi_y \leq b$, where $b$ is the inter-chain distance
and with $\nu \propto 1/v_F b$, one has: $T_0 \propto 1 / \tau_{el}$,
up to a numerical factor.
\bibitem{Note:ls}
This follows from the fact that is quasi-1D systems, charges are constrained
to move along lines, with a small hopping amplitude from one line to the
other. For the Wigner crystal, where displacements are two-dimensional,
the basic defect is the vacancy or the interstitial, with a length
of the order of the distance between electrons.
\bibitem{Note:fscatt}
The f-scattering term is generally removed by a shift of the $\varphi-$field. One then works on the background of a deformed system. Here, we are interested in such deformations and we therefore consider explicitly the f-scattering term.
\end{thebibliography}
\end{document}